\documentclass[a4paper,11pt]{article}
\pdfoutput=1 

\usepackage{jcappub} 

\usepackage[T1]{fontenc} 
\usepackage{xcolor}
\usepackage{threeparttable}

\title{The role of magnetospheric current sheets in pair enrichment and ultra-high energy proton acceleration in M87*}

\author[a,b]{S.~I. Stathopoulos,\note{Corresponding author.}}
\author[a,b]{M. Petropoulou,}
\author[c,d]{L. Sironi,}
\author[e]{D. Giannios}

\affiliation[a]{Department of Physics, National and Kapodistrian University of Athens, University Campus Zografos, GR 15783, Greece}
\affiliation[b]{Institute of Accelerating Systems \& Applications, University Campus Zografos, Athens, Greece}
\affiliation[c]{Department of Astronomy and Columbia Astrophysics 
Laboratory, Columbia University, New York, NY 10027, USA}
\affiliation[d]{Center for Computational Astrophysics, Flatiron Institute, 162 5th Avenue, New York, NY 10010, USA}
\affiliation[e]{Department of Physics, Purdue University, West Lafayette, IN, 47907, USA}
\emailAdd{stamstath@phys.uoa.gr}
\emailAdd{mpetropo@phys.uoa.gr}
\emailAdd{lsironi@astro.columbia.edu}
\emailAdd{dgiannio@purdue.edu}

\abstract{Recent advances in numerical simulations of magnetically arrested accretion onto supermassive black holes have shed light on the formation and dynamics of magnetospheric current sheets near the black hole horizon. By considering the pair magnetization $\sigma_{\rm e}$ in the upstream region and the mass accretion rate $\dot{m}$ (in units of the Eddington mass accretion rate) as free parameters we estimate the strength of the magnetic field and develop analytical models, motivated by recent three-dimensional particle-in-cell simulations, to describe the populations of relativistic electron and positrons (pairs) in the reconnection region.  
Applying our model to M87*, we numerically compute the non-thermal photon spectra for various values of $\sigma_e$. We show that pairs that are accelerated up to the synchrotron radiation-limited energy while meandering across both sides of the current sheet, can produce MeV flares with luminosity of $\sim 10^{41}$~erg s$^{-1}$ -- independent of $\sigma_e$ -- for a black hole accreting at $\dot{m}=10^{-5}$. Pairs that are trapped in the transient current sheet can produce X-ray counterparts to the MeV flares, lasting about a day for current sheets with length of a few gravitational radii. We also show that the upstream plasma can be enriched due to photon-photon pair creation, and derive a new equilibrium magnetization of $\sigma_e \sim 10^3-10^4$ for $\dot{m}=10^{-6} - 10^{-5}$. Additionally, we explore the potential of magnetospheric current sheets to accelerate protons to ultra-high energies, finding that while acceleration to such energies is limited by various loss mechanisms, such as synchrotron and photopion losses from the non-thermal emission from pairs, maximal proton energies in the range of a few EeV are attainable in magnetospheric sheets forming around supermassive sub-Eddington accreting black holes.}

\begin{document} 
\maketitle
\flushbottom

\section{Introduction}
An accreting supermassive black hole (SMBH) is the powerhouse of an active galactic nucleus (AGN). M87 is a massive elliptical radio galaxy at a distance of $16.8$ Mpc and has a black hole mass about $6.5\cdot 10^9 M_{\odot }$ \cite{Akiyama_2019}. The non-thermal electromagnetic emission emerging from the vicinity of its SMBH, M87*, is variable,  occasionally exhibiting flares in X-rays and very high-energy (VHE, $>0.35$~TeV) $\gamma$-rays \cite{2009Sci...325..444A,Harris_2009,Aliu_2012}, on timescales as short as a few times the light crossing time ($1-20t_{\rm cr}$) of the black hole's gravitational radius. The production of non-thermal radiation from the vicinity of the SMBH conveys the potential for particle acceleration to high energies, thereby suggesting the plausibility of proton acceleration to ultra-high energies. Notably, M87* has been postulated as a potential accelerator of Ultra High Energy Cosmic Rays (UHECRs) \cite{2016Natur.531..476H,Rieger_2022}. The properties of the accelerated particle population, such as their distribution in energy and maximal energy reached, are determined by the dominant particle acceleration mechanism at work and the physical conditions in the accelerator region. Recently, the Event Horizon Telescope (EHT) has provided insights into the innermost regions of the accretion flow surrounding M87*. By analyzing these images alongside predictions from general relativistic magnetohydrodynamic (GRMHD) models, the Collaboration suggests a scenario in which a magnetically arrested disk (MAD) exists around a black hole rotating with a moderate to high spin parameter \cite{Akiyama_2021}.
 
High-resolution three-dimensional (3D) GRMHD simulations of accretion onto rotating SMBHs have revealed transient and non-axisymmetric magnetospheres during periods of magnetic flux decay at the horizon, leading to significant drops in mass accretion rates and the emergence of thin equatorial current sheets separating the two sides of the polar jet \cite{2022ApJ...924L..32R}. The intricate dynamics of these current sheets, including plasmoid-mediated reconnection and the injection of reconnection-heated plasma into the accretion disk and jet boundary, can provide a dynamic environment where particles can accelerate and eventually power outbursts of electromagnetic radiation, commonly known as flares. However, challenges persist in understanding the triggering mechanisms behind large flux eruption events and how they are impacted by the system's physical parameters, such as the spin of the SMBH, or when the system is not in the MAD regime, and others. 

Magnetic reconnection in plasmas with high magnetization ($\sigma \gg 1$), defined as twice the ratio of the magnetic energy density to the plasma enthalpy density, is an efficient process of magnetic energy dissipation. A significant fraction of the dissipated energy is used to accelerate particles to relativistic energies $\sim \sigma m c^2$ \cite{Lyutikov_2003,Lyubarsky_2005,2016ApJ...816L...8W}. The physics of reconnection can only be captured from first principles utilizing fully kinetic particle-in-cell (PIC) simulations. A large body of work on 2D simulations of relativistic reconnection in pair plasmas has demonstrated that particles are accelerated into power-law distributions, $dN/d\gamma \propto \gamma^{-p}$, that extend up to $\sim \sigma$ and with a power-law slope $p$ that depends on the plasma magnetization \cite{Sironi_2014,2016ApJ...816L...8W}. Particles reaching the highest energies were injected into the acceleration process by interactions with the non-ideal electric fields at the so-called X-points of the current sheet, regions where the conditions of ideal MHD break down \cite{2022PhRvL.128n5102S}. A secondary acceleration that could push particles to $\gamma \gg \sigma$ was also found to operate for particles trapped in compressing plasmoids, albeit on much longer timescales \cite{2018MNRAS.481.5687P, 2021ApJ...912...48H}. Recent large-box 3D PIC simulations of reconnection in pair plasmas have revealed that particles with $\gamma>\sigma$ can escape from plasmoids (flux ropes in 3D) and can accelerate to even higher energies while meandering between the two sides of the reconnection layer in the upstream (inflow) region \cite{Zhang_2021, 2023ApJ...956L..36Z}. These particles will terminate their acceleration phase by reentering into the flux ropes where the acceleration is not efficient or escape from the system.

The primary goal of this paper is to examine the role of magnetospheric current sheets in the context of magnetically arrested accretion onto SMBHs, with a particular focus on  M87*. We explore the evolution of relativistic particle populations accelerated in current sheets during magnetic reconnection, and investigate their implications for the production of non-thermal radiation and the potential acceleration of ultra-high energy protons. To achieve this goal, we employ a two-step approach that combines theoretical modeling and numerical calculations. Motivated by the findings of recent 3D PIC simulations, we develop a model for non-thermal emission from the reconnection region, accounting for the presence of two distinct particle populations (``free'' and ``trapped'', located at the reconnection upstream and flux ropes, respectively) \cite{2023ApJ...956L..36Z}, as well as the interplay between these two populations and secondary pairs injected into the system through $\gamma \gamma$ pair creation. The main parameters of the model are the pair magnetization in the upstream region of the current sheet, the length of the current sheet, and the mass accretion rate. Subsequently, we utilize numerical methods to calculate the non-thermal radiation spectra arising from these pair populations within a one-zone leptonic model, considering interactions with both non-thermal and disk photon fields. By applying our model to the specific case of M87*, we aim to provide insights into the observed phenomena and discuss the implications of our findings on magnetospheric current sheets and their potential role in particle acceleration near SMBHs.
    
This paper is structured as follows. In Sec. \ref{sec:Th_fr} we outline our theoretical framework and present the key parameters of our model. In Sec. \ref{sec:free} we describe the pairs that are in the free acceleration phase while in Sec. \ref{sec:trap} we discuss the pair dynamics in the downstream region analytically. In Sec. \ref{sub_sec:Numerical_appr} we showcase the numerical approach to the problem. In Sec. \ref{sec:m87} we apply our model to M87*. In Sec. \ref{sec:spectra} we present the numerical results of the pair and photon spectrum. Using these numerical results in Sec. \ref{sub_sec:con_f_obv} we determine which combinations of the key parameters can reproduce some of the observations in the X-rays. In Sec. \ref{sub_sec:UHECR} we check whether the production of ultra-high energy protons in M87* is feasible. In Sec. \ref{sub_sec:pair_enr} we discuss what is the effect of the pair creation in the system. Finally, we present the conclusions of this work in Sec. \ref{sec:discuss}.

\section{Theoretical framework}\label{sec:Th_fr}

Recent 3D high-resolution general relativistic MHD simulations of accretion onto a black hole through a magnetically arrested disk \citep[MAD,][]{2003PASJ...55L..69N, 2011MNRAS.418L..79T} have revealed the formation of short-lived current sheets in the black hole magnetospheric region \citep{2022ApJ...924L..32R}. The reconnection layers have a typical length of a few gravitational radii of the supermassive black hole (SMBH), i.e. $l \lesssim 10 r_g$, where $r_g = GM/c^2 \simeq 1.6 \cdot 10^{14}~M/(10^9 M_\odot)$~cm, and their lifespan is of the order of $10 \ l/c$ for the largest current sheets \cite{2022ApJ...924L..32R}.

One can estimate the strength of the magnetic field threading the black hole horizon by noting that in the MAD regime the dimensionless magnetic flux threading the black hole, $\phi_{\rm BH}=\Phi_{\rm BH}/\sqrt{\dot{M} cr_{\rm g}^2}$, cannot exceed an approximate value of 50 \citep{2011MNRAS.418L..79T}. This translates to the relation

\begin{equation}
\Phi_{\rm BH}\sim 50\sqrt{ \dot{M} c r_{\rm g}^2},
\label{eq:magn_flux_MAD}
\end{equation}
where $\dot{M}$ is the accretion rate onto the SMBH, and $\Phi_{\rm BH} = 4 \pi r_{\rm H}^2 B_0$ is the magnetic flux threading the black hole horizon of radius $r_{\rm H}=r_{\rm g} f(a_{\rm s})$, where $f(a_{\rm s})=(1+\sqrt{1-a^2_{\rm s}})$ and $a_s$ is the dimensionless spin of the SMBH. Introducing the dimensionless accretion rate, $\dot{m} = \dot{M}/\dot{M}_{\rm Edd}$, where $\dot{M}_{\rm Edd} = L_{\rm Edd}/(\eta_{\rm c} c^2)$ is the Eddington accretion rate and $\eta_{\rm c}$ is a matter-to-luminosity conversion factor \cite{2002apa..book.....F}, we can estimate the magnetic field strength as

\begin{equation}
B_0\simeq 5 \cdot 10^2~{\rm G} \, \frac{\dot{m}_{-5}^{1/2}(M_9\eta_{\rm c,-1})^{-1/2}}{f^2(a_{\rm s})}
\label{eq:magn_field}
\end{equation}
where we have introduced the notation $Q_{\rm x} = Q/10^x$ and $M$ is expressed in units of the solar mass. In what follows, we assume that $B_0$ is the typical magnetic field strength in the upstream region of the magnetospheric current sheets.

Recent 3D kinetic simulations of reconnection in pair plasmas \citep{2021ApJ...922..261Z, 2023ApJ...956L..36Z} have shown that the total pair distribution in the reconnection region is composed of the so-called ``free'' particles that undergo active acceleration in the inflow region, by meandering between the two sides of the reconnection layer, and the ``trapped'' particles that end up inside plasmoids (flux tubes) and do not undergo further acceleration. These can cool due to radiative losses and eventually get advected out from the reconnection region. We develop an analytical model to describe the steady-state populations of free and trapped particles and then complement our analysis with numerical calculations of non-thermal radiation accounting for physical processes, like inverse Compton scattering and photon-photon pair production.

We assume that cold pairs dominate the upstream plasma. The pair magnetization $\sigma_{\rm e}$, which is one of the main free parameters of the model, is then defined as
\begin{equation}
\sigma_{\rm e} = \frac{B_0^2}{4\pi n_{\rm e^\pm} m_{\rm e} c^2},
\label{eq:sigma}
\end{equation}
where $n_{\rm e^\pm}$ is the pair density in the upstream. The injection rate of pairs to the downstream region can be written as,
\begin{equation} 
Q^{\rm tot}_{\rm e,inj}=2 n_{\rm e^{\pm}} \eta_{\rm rec}c A.
\label{eq:Inj_rate_up}
\end{equation}
where $\eta_{\rm rec}=v_{\rm rec}/c\sim0.06$ is the quasi-steady state reconnection rate achieved in 3D PIC simulations of reconnection \citep{2021ApJ...922..261Z,2023ApJ...956L..36Z}, $A = \pi R^2$ is the area of a cylindrical current sheet of radius $R\equiv \mathcal{R}r_g$, and the factor of 2 accounts for the inflow of particles from both sides of the sheet. We assume that a fraction $\zeta$ of the pairs entering the layer will become ``free'' after being first accelerated to a Lorentz factor $\gamma_{\rm inj} \sim \sigma_{\rm e} \gg 1$ within the reconnection layer. The free pairs may leave the upstream region of active acceleration and become eventually trapped in the plasmoids of the current sheet, where they continue cooling due to radiative losses. The remaining  $(1-\zeta)$ fraction of the pairs entering the layer will only experience the injection stage
within the current sheet (e.g. at X-points \cite{2022PhRvL.128n5102S, 2023PhRvL.130r9501G}), without ever becoming free in the upstream, before getting trapped in plasmoids. We adopt $\zeta = 0.06$ as a typical value that is motivated by simulations \cite{2023ApJ...956L..36Z} (for more details, see Appendix~\ref{app:zeta}).

In what follows we compute the distribution functions of the free and trapped pairs, and the resulting synchrotron spectra using a semi-analytical approach.

\subsection{The ``free'' pair distribution}\label{sec:free}
We first consider the emission produced by pairs in the free acceleration phase. In the free phase pairs can be accelerated by the ideal electric field and lose energy due to synchrotron radiation\footnote{In the analytical model we ignore inverse Compton losses (ICS) due to disk radiation, but these are included in the numerical calculations. Nevertheless, the magnetic energy density is typically much larger than the disk photon radiation density, making ICS losses negligible. For instance, using parameters relevant to M87*, we find 
\begin{equation}
\frac{U_{\rm B}}{U_{\rm soft}} = \frac{B_0^2/(8\pi)}{3 L_{\rm soft}/(4\pi R_{\rm soft}^2 c)} \simeq 10^{4},
\end{equation}
where $L_{\rm soft} \approx \nu_{\rm EHT} L_{\nu_{\rm EHT}} \simeq 10^{41} $~erg s$^{-1}$, $\nu_{\rm EHT} =230$~GHz, $R_{\rm soft} = 5 r_{\rm g}$ and, $B_0=100G$.}. The kinetic equation governing their differential number distribution, $N^{\rm free}(\gamma)\equiv \mathrm{d}N^{\rm free}/ \mathrm{d}\gamma$, is written as~\cite{1998A&A...333..452K}:

\begin{equation}
\frac{\partial N^{\rm free}}{\partial t}+\frac{\partial }{\partial \gamma}\Big( (\dot{\gamma}_{\rm acc}+\dot{\gamma}_{\rm syn})N^{\rm free}\Big)+\frac{N^{\rm free}}{t^{\rm fr}_{\rm esc}(\gamma)} = \zeta Q^{\rm tot}_{\rm e, inj} \delta(\gamma-\gamma_{\rm inj}),
\label{eq:kin_eq_fp}
\end{equation}
where $\gamma_{\rm inj} \sim \sigma_{\rm e}$. Using Eqs.~(\ref{eq:magn_field}) and (\ref{eq:sigma}) the free particle injection rate can be expressed in terms of the main model parameters as
\begin{equation}
Q^{\rm free}_{\rm e,inj} = \zeta Q^{\rm tot}_{\rm e,inj} 
\simeq  7\cdot 10^{47}~{\rm s}^{-1}~\frac{\zeta_{-1.2}}{\sigma_{\rm e}}\frac{\eta_{\rm rec,-1}M_9\dot{m}_{-5} \mathcal{R}_0 ^2}{f^4(a_{\rm s})}.
\label{eq:Q_inj}
\end{equation}
The corresponding energy injection rate can be estimated as
\begin{equation}
    L^{\rm free}_{\rm e, inj} \approx Q^{\rm free}_{\rm e, inj} \sigma_{\rm e} m_{\rm e} c^2 \simeq 0.2 \ \zeta_{-1.5} \ \eta_{\rm rec, -1} \, L_{\rm BZ},
    \label{eq:Leinj}
\end{equation}
where $L_{\rm BZ} \simeq \pi r_{\rm g}^2 B_0^2 c/24$ is the power extracted from a maximally rotating black hole through the Blandford - Znajek process \citep{1977MNRAS.179..433B, 2011MNRAS.418L..79T}.

The acceleration and synchrotron loss rates in Eq.~(\ref{eq:kin_eq_fp}) are defined respectively as
\begin{equation}
\dot{\gamma}_{\rm acc}= \frac{\eta_{\rm rec}eB_0 \beta_{\rm z}}{m_{\rm e} c}\equiv \beta_{\rm a},
\label{eq:dot_acc}
\end{equation}
and
\begin{equation}
\dot{\gamma}_{\rm syn}= -\frac{\sigma_{\rm T}B_0^2}{6\pi m_{\rm e} c} \gamma^2 \equiv -\beta_{\rm s} \gamma^2,
\label{eq:dot_syn}
\end{equation}
where $\beta_{\rm z} \sim 1$ is the particle velocity along the $z$ direction of the electric current in units of $c$ (in what follows we set $\beta_z=1$ and do not show it explicitly). We note that a pitch angle of $\pi/2$ was assumed in Eq.~(\ref{eq:dot_syn}) because free particles move preferentially along the $z$ direction and therefore perpendicular to the magnetic field direction. Finally, $t^{\rm fr}_{\rm esc}(\gamma)$ is the escape timescale from the free acceleration phase, 
\begin{equation}
t^{\rm fr}_{\rm esc} \simeq t_{\rm acc} \equiv \gamma/\dot{\gamma}_{\rm acc}=6.3\cdot 10^{-4}~{\rm s} \, \gamma \,  \eta_{\rm rec,-1}^{-1}(\sigma_{\rm e}n_{\rm e^{\pm}})^{-1/2},
\label{eq:t_esc}
\end{equation}
as found in PIC simulations \cite{2023ApJ...956L..36Z}. 

The solution of Eq.~(\ref{eq:kin_eq_fp}) is~\citep{1998A&A...333..452K, 2023ApJ...956L..36Z}:

\begin{equation}
N^{\rm free}(\gamma,t) = \frac{Q^{\rm free}_{\rm e, inj}\gamma_{\rm rad}^2}{\beta_{\rm a}(\gamma_{\rm rad}^2-\gamma^2)}\bigg[\frac{\gamma}{\gamma_{\rm inj}}\sqrt{\frac{\gamma_{\rm rad}^2-\gamma_{\rm inj}^2}{\gamma_{\rm rad}^2-\gamma^2}}\bigg]^{-s_{\rm free}},  t \ge \tau(\gamma),
\label{eq:kin_eq_fp_sol}
\end{equation}
where $s_{\rm free}=t_{\rm acc}/t_{\rm esc} = 1$, and the characteristic timescale $\tau$ is given by

\begin{equation}
\tau(\gamma)=\frac{\gamma_{\rm rad}}{2\beta_{\rm a}}\ln\bigg[\frac{(\gamma_{\rm rad}+\gamma)(\gamma_{\rm rad}-\gamma_{\rm inj})}{(\gamma_{\rm rad}-\gamma)(\gamma_{\rm rad}+\gamma_{\rm inj})}\bigg].
\label{eq:tau_gg}
\end{equation}

In the above expressions $\gamma_{\rm rad}=\sqrt{\beta_{\rm a}/\beta_{\rm s}}$ is the synchrotron radiation-limited Lorentz factor of pairs, which is found by solving $\dot{\gamma}_{\rm acc} = -\dot{\gamma}_{\rm syn}$, and it can also be expressed as

\begin{equation} 
\gamma_{\rm rad} \simeq \sqrt{3}\cdot 10^6 \
M_9^{1/4} \eta_{\rm c,-1}^{-1/4} \dot{m}_{-5}^{-1/4} \eta_{\rm rec, -1}^{1/2} f(a_{\rm s}).
\label{eq:gamma_rad}
\end{equation}

From this point on we restrict our analysis to the regime of $\sigma_{\rm e} < \gamma_{\rm rad} \sim 10^6$ because the free particle channel would not exist otherwise. We will return to this point later in the Discussion. 

The pair distribution of Eq.~(\ref{eq:kin_eq_fp_sol}) can be approximated by a power law for $\gamma \ll \gamma_{\rm rad}$,
\begin{equation}
N^{\rm free}(\gamma)\simeq \frac{Q^{\rm free}_{\rm e,inj}}{\beta_{\rm a}}\bigg(\frac{\gamma}{\gamma_{\rm inj}}\bigg)^{-1}, \, \gamma_{\rm inj} \le \gamma \ll \gamma_{\rm rad}.
\label{eq:pair_distribution_approx}
\end{equation} 
and the total energy of the free population reads 

\begin{eqnarray}    
\label{eq:E_free}
E_{\rm free} &=& m_{\rm e} c^2 \int_{\gamma_{\rm inj}}^{\gamma_{\rm rad}} d\gamma \ \gamma  N^{\rm free}(\gamma) 
\simeq  \frac{Q^{\rm free}_{\rm e,inj}}{\beta_{\rm a}} \gamma_{\rm inj} m_{\rm e} c^2 \gamma_{\rm rad}
\simeq \frac{A}{2\pi} \zeta m_{\rm e} c^2 \left(\frac{6 \pi \eta_{\rm rec} B_0}{e \sigma_{\rm T}}\right)^{1/2}\simeq  \nonumber \\ 
&\simeq& 10^{39}~{\rm erg} \, \zeta_{-1.2}\mathcal{R}_0^2 M_9^{7/4}\dot{m}_{-5}^{1/4}\eta_{\rm c,-1}^{-1/4}\eta_{\rm rec,-1}^{1/2}f(a_{\rm s})^{-1},
\end{eqnarray}
where we have assumed that $\gamma_{\rm inj} \ll \gamma_{\rm rad}$.

Meanwhile, the total available energy in the system (i.e., the dissipated magnetic energy in the current sheet's lifetime) can be estimated as
\begin{equation}
E_{\rm tot} \simeq 2\frac{c}{4\pi} E_{\rm rec} B_0 \, A \, T \simeq 30\frac{\eta_{\rm rec} B_0^2}{4 \pi} A \, R ,
\end{equation}
where $T = 15 (R/c) \simeq 0.9~\mathcal{R} M_9$~days is the assumed lifetime of the current sheet, and $E_{\rm rec} = \eta_{\rm rec} B_0$ is the ideal electric field in the upstream region of the current sheet. The energy ratio is then given by
\begin{equation}
\frac{E_{\rm free}}{E_{\rm tot}} \simeq  \frac{\zeta  m_{\rm e} c^2}{15 R} \left( \frac{6\pi}{e \sigma_{\rm T} \eta_{\rm rec} B_0^3}\right)^{1/2} 
\simeq 5 \cdot 10^{-10} \zeta_{-1.2} \eta_{\rm rec,-1}^{-1/2} \dot{m}_{-5}^{-3/4} \eta_{\rm c, -1}^{3/4} M_9^{-1/4}\mathcal{R}_0^{-1} f^3(a_{\rm s}),
\label{eq:energy-ratio}
\end{equation}
where we used Eq.~(\ref{eq:magn_field}). Only a small fraction of the available system energy is carried by the free particles, as their maximal energy is strongly limited by the synchrotron losses and they cannot tap all the available energy. This can also be verified by comparing $\gamma_{\rm rad}$ in Eq.~(\ref{eq:gamma_rad}) with the maximum Lorentz factor attained within the acceleration region of length $l=2R$ and electric field $E_{\rm rec}$,
\begin{equation}
\gamma_{\rm e, \max} \simeq \frac{e \eta_{\rm rec} B_0 l}{m_{\rm e} c^2}\simeq 8 \cdot 10^{12}\frac{\eta_{\rm rec,-1}(\dot{m}_{-5}M_9)^{1/2}\mathcal{R}_0\eta_{\rm c, -1}^{-1/2}}{f^2(a_{\rm s})}.
\label{eq:gamma_e_max}
\end{equation}

Using Eqs.~(\ref{eq:gamma_rad}) and (\ref{eq:gamma_e_max}), the energy ratio of Eq.~(\ref{eq:energy-ratio}) can be simply expressed as
\begin{equation}
\frac{E_{\rm free}}{E_{\rm tot}} \simeq \frac{2\zeta}{15}\frac{\gamma_{\rm rad}}{\gamma_{\rm e,\max}} \ll 1.
\end{equation}

Because of the hard power-law distribution of free pairs ($s_{\rm free} = 1)$ their synchrotron spectrum will peak roughly at the photon energy corresponding to the burnoff limit

\begin{equation}
\epsilon_{\rm syn} \simeq \frac{B_0}{B_{\rm cr}}\gamma_{\rm rad}^2 m_{\rm e} c^2 \simeq
10~{\rm MeV} \  \eta_{\rm rec,-1},  
\label{eq:nu_c}
\end{equation}
where $B_{\rm cr} = m_{\rm e}^2 c^2 / (e \hbar) \simeq 4.4\cdot 10^{13}$~G is the Schwinger magnetic field strength. The corresponding peak synchrotron luminosity at the target photon energy can be estimated as 
\begin{equation}
L^{\rm pk}_{\rm syn}\approx \frac{\sigma_{\rm T} c}{6\pi} B_0^2 \gamma^2_{\rm rad}  \Bigg[\gamma \frac{dN^{\rm free}} {d\gamma} \Bigg]_{\gamma=\gamma_{\rm rad}} \!\!\!\!\! \!\!\!\!\!  \simeq L^{\rm free}_{\rm e, inj} \simeq 6\cdot 10^{41}~{\rm erg \ s^{-1}} \zeta_{-1.2} \ \eta_{\rm rec,-1} M_9 \dot{m}_{-5} \frac{\mathcal{R}_0^2}{f^{4}(a_{\rm s})},
\label{eq:nLn_syn}
\end{equation}
where we used Eqs.~(\ref{eq:Q_inj}) and (\ref{eq:Leinj}). The synchrotron luminosity of the free pairs and the peak synchrotron photon energy are independent of the uncertain pair magnetization in the magnetospheric region.

The peak synchrotron photons may pair produce (close to the peak of the cross-section) on less energetic synchrotron photons with energy

\begin{equation}
\epsilon_{\rm l} \approx 2 \frac{(m_{\rm e} c^2)^2}{\epsilon_{\rm syn}} \approx 5 \cdot 10^{-2}~{\rm MeV} \, \eta_{\rm rec,-1}^{-1}.
\label{eq:epsilon_l}
\end{equation}
The corresponding synchrotron luminosity is

\begin{equation}
L_{\rm l} \approx L^{\rm pk}_{\rm syn}  \left(\frac{\epsilon_{\rm l}}{\epsilon_{\rm syn}}\right)^{4/3} \simeq 8.5 \cdot 10^{-4} \,  L^{\rm pk}_{\rm syn} \, \eta_{\rm rec,-1}^{-8/3}.
\label{eq:L_l}
\end{equation}
The optical depth for the attenuation of the peak synchrotron photons can be approximated as 
\begin{equation}
\tau_{\rm \gamma \gamma}(\epsilon_{\rm syn}) \approx \frac{\sigma_{\rm T}}{4} n_{\rm l}  R \simeq \frac{3 \sigma_{\rm T} L_{\rm l} } {16 \pi R c \epsilon_{\rm l}} \simeq 9 \cdot  10^{-5} \eta_{\rm rec,-1}^{-5/3} L^{\rm pk}_{\rm syn, 42} \mathcal{R}^{-1}_0,
\label{eq:tau_gg_free}
\end{equation}
where $n_{\rm l}$ is the number density of target photons of energy $\epsilon_{\rm l}$, and $\sigma_{\rm T}/4$ is the approximate peak value for the cross section \cite{1990MNRAS.245..453C}. The energy injected per unit time into secondary pairs from $\gamma \gamma$ production can be then written as

\begin{equation}
L_{\rm inj}^{\rm sec}\approx \tau_{\gamma \gamma}(\epsilon_{\rm syn}) L^{\rm pk}_{\rm syn} \simeq 9  \cdot 10^{37}~{\rm erg \ s^{-1}} (L^{\rm pk}_{\rm syn, 42} )^2 \eta_{\rm rec,-1}^{-5/3} \mathcal{R}^{-1}_0,
\label{eq:L_sec}
\end{equation}
which is orders of magnitude lower than the free pair injection luminosity, is independent of $\sigma_{\rm e}$, and scales quadratically with $\dot{m}$. Secondary pairs will be injected into the system with 
\begin{equation} 
\gamma_{\rm inj}^{\rm sec}\approx \frac{\epsilon_{\rm syn}}{2 m_{\rm e} c^2} \simeq 10~\eta_{\rm rec,-1},
\label{eq:gamma_sec}
\end{equation}
and their steady state distribution can be obtained from Eq.~(\ref{eq:kin_eq_fp}) after using the appropriate source term ($Q_{\rm inj}^{\rm sec}\approx 2 \tau_{\rm \gamma \gamma} L^{\rm pk}_{\rm syn}/\epsilon_{\rm syn}$), and neglecting the acceleration term. Therefore, $\gamma \gamma$ pair production is not expected to alter the total pair density in the system or impact the overall photon spectrum, unless more low-energy photons are available, in addition to the ones provided by the free particles (as we assumed so far). In the next section, we will investigate whether the synchrotron radiation of trapped pairs may provide these additional target photons.
 
\subsection{The ``trapped'' pair distribution}\label{sec:trap}
Trapped pairs can be classified into two distinct groups. The first group consists of free pairs that escape the (upstream) region of active acceleration and become trapped in the plasmoids of the reconnection region. These pairs are injected into the trapped phase with Lorentz factor from $\gamma_{\rm inj}\sim \sigma_{\rm e}$ up to $\gamma_{\rm rad}$, but with a softer power law than the free pairs, as we describe below. The second group of trapped pairs consists of particles that were accelerated in the current sheet up to $\sigma_{\rm e}$ (at e.g. X-points \cite{2022PhRvL.128n5102S}) and then became trapped in plasmoids where no further fast acceleration takes place. Consequently, we can express the total injection rate to the trapped population (integrated over $\gamma$) as follows,

\begin{equation}
Q^{\rm trap}_{\rm e}=Q^{\rm trap}_{\rm X}+Q^{\rm trap}_{\rm fr},
\label{eq:trapped_pairs_rate}
\end{equation}
where $Q^{\rm trap}_{\rm X}$ corresponds to the injection rate of trapped particles that went through the injection phase of acceleration but did not experience a free phase of acceleration, and $Q^{\rm trap}_{\rm fr}$ indicates the injection rate of pairs from the free to the trapped phase.

\subsubsection{The free-trapped channel}
The differential injection rate of pairs from the free to the trapped phase can be expressed as
\begin{equation}
q^{\rm trap}_{\rm fr}(\gamma>\gamma_{\rm inj}) \equiv \frac{dN^{\rm trap}_{\rm fr}}{d\gamma dt}\simeq \frac{1}{t^{\rm fr}_{\rm esc}(\gamma)}\frac{dN^{\rm free}}{d\gamma} \simeq Q^{\rm free}_{\rm e,inj}\gamma_{\rm inj}\gamma^{-2},
\label{eq::trapped_pairs_approx}
\end{equation}
where we used Eq.~(\ref{eq:pair_distribution_approx}) to derive the right-hand side of the expression. These pairs that are trapped in the reconnection region may escape on the advection timescale of plasmoids, i.e. $t_{\rm adv} = R/V_{\rm A} \sim R/c$, where $V_{\rm A} = c\sqrt{\sigma_{\rm e}/(1+\sigma_{\rm e})}$ is the Alfv{\'e}n speed~(see also Appendix B of Ref.~\cite{2021ApJ...922..261Z}). Meanwhile, they experience synchrotron losses in a magnetic field with comparable strength to the upstream region ~\citep{10.1093/mnras/stw1620}. 

The steady-state trapped pair distribution originating from the free phase can be obtained by solving Eq.~(\ref{eq:kin_eq_fp}) without the acceleration term, after using Eq.~(\ref{eq::trapped_pairs_approx}) as the injection term\footnote{In the analytical model, we neglect the injection of secondary pairs below $\gamma_{\rm inj}$ from $\gamma \gamma$ pair production. Their contribution will be assessed in the next section through numerical calculations.}, and after replacing $t^{\rm fr}_{\rm esc}(\gamma)$ with $t_{\rm adv}$ (see also Ref.~\cite{2023ApJ...956L..36Z}),

\begin{equation}
N^{\rm trap}_{\rm fr}(\gamma) \simeq  3\cdot 10^{49}~\frac{\zeta_{-1.2}\eta_{\rm rec,-1}M_9^2\dot{m}_{-6} \mathcal{R}_0 ^3}{f^4(a_{\rm s})}\gamma^{-2}\cdot
\begin{cases} 
      1-e^{\frac{1}{\beta_{\rm s} t_{\rm adv}}(\frac{1}{\gamma_{\rm rad}}-\frac{1}{\gamma})}, & \gamma_{\rm inj} < \gamma \le \gamma_{\rm rad}\\
      e^{\frac{1}{\beta_{\rm s} t_{\rm adv}}(\frac{1}{\gamma_{\rm inj}}-\frac{1}{\gamma})}-e^{\frac{1}{\beta_{\rm s} t_{\rm adv}}(\frac{1}{\gamma_{\rm rad}}-\frac{1}{\gamma})}, & 1<\gamma \le \gamma_{\rm inj}\\
   \end{cases}
\label{eq:trapped_pairs_dis}
\end{equation}
The second branch of the expression above is relevant when the Lorentz factor $\gamma^{\rm syn}_{\rm cool}$ of an electron that cools on a dynamical time scale is $\gamma^{\rm syn}_{\rm cool} \ll \gamma_{\rm inj}$, where $\gamma^{\rm syn}_{\rm cool}$ is given by

\begin{equation}
    \gamma^{\rm syn}_{\rm cool} = \frac{6 \pi m_{\rm e} c^2 }{B_0^2 \sigma_{\rm T} R} = 2\frac{\gamma_{\rm rad}^2}{\gamma_{\rm e,\max}}.
    \label{eq:syn_cool}
\end{equation}

The condition for fast cooling, $\gamma^{\rm syn}_{\rm cool} \ll \gamma_{\rm inj}$, translates to $2 \ll \sigma_{\rm e}\eta_{\rm c,-1}(\dot{m}_{-5}\mathcal{R}_0)^{-1}$. If $\beta_{\rm s} t_{\rm adv} \gamma\gg 1$, or equivalently, if $t_{\rm adv}/t_{\rm syn} \gg 1$ (where we used Eq.~\ref{eq:dot_syn}), the steady-state distribution can be approximated by a broken power law,

\begin{equation}
N^{\rm trap}_{\rm fr}(\gamma) \simeq 2\cdot  10^{50}~\zeta_{-1.2}\eta_{\rm rec,-1}\eta_{\rm c,-1}M_9^2 \mathcal{R}_0^2\gamma^{-2}\cdot
\begin{cases} 
      \frac{1}{\gamma}-\frac{1}{\gamma_{\rm rad}}, & \gamma_{\rm inj} < \gamma \le \gamma_{\rm rad} \\
      \frac{1}{\gamma_{\rm inj}}-\frac{1}{\gamma_{\rm rad}}, & 1<\gamma\le \gamma_{\rm inj}\\
   \end{cases}
\label{eq:trapped_pairs_dis_approx}
\end{equation}
The total number of trapped particles originating from the free phase is given by 
\begin{eqnarray} 
N^{\rm trap}_{\rm fr,tot} &=& \int_1^{\gamma_{\rm rad}} \!\!\! \! \! d\gamma N^{\rm trap}_{\rm fr}(\gamma)  \simeq 2\cdot  10^{50}~\zeta_{-1.2}\eta_{\rm rec,-1}\eta_{\rm c,-1}M_9^2 \mathcal{R}_0^2 \Bigg(\frac{1}{2\gamma_{\rm rad}^2}-\frac{1}{2\gamma_{\rm inj}^2}+\frac{1}{\gamma_{\rm inj}}-\frac{1}{\gamma_{\rm rad}}\Bigg) \nonumber  \\ 
&\approx&  2\cdot  10^{50}~\zeta_{-1.2}\eta_{\rm rec,-1}\eta_{\rm c,-1}M_9^2 \mathcal{R}_0^2 \gamma^{-1}_{\rm inj},
\label{eq:trapped_pairs_num_approx}
\end{eqnarray}  
where the expression in the second line has been derived in the limit of $1 \ll \gamma_{\rm inj} \ll \gamma_{\rm rad}$. The total number of trapped particles originating from the free phase scales as $\gamma_{\rm inj}^{-1} \propto \sigma_{\rm e}^{-1}$, since it is determined by the number of free pairs whose injection rate scales also as $\sigma_{\rm e}^{-1}$ (see Eq. \ref{eq:Q_inj}). In this limit, the total energy of the trapped pair population reads,
\begin{eqnarray} 
E^{\rm trap}_{\rm fr,tot} \simeq m_{\rm e} c^2 \int_1^{\gamma_{\rm rad}} \!\!\! \! \! d\gamma \gamma N^{\rm trap}_{\rm fr}(\gamma)  = 8 \cdot 10^{43}~{\rm erg}~\zeta_{-1.2}\eta_{\rm rec,-1}\eta_{\rm c,-1}M_9^2 \mathcal{R}_0^2\frac{1+\ln(\gamma_{\rm inj})}{\gamma_{\rm inj}} ,
\label{eq:E_fr_trapped_pairs_approx}
\end{eqnarray}  
and is much larger than the energy of the free particle population (see Eq.~\ref{eq:E_free}).

\subsubsection{The X-trapped channel}
The rate at which pairs are injected directly into the e.g., X-point phase of acceleration is parameterized as $(1-\zeta) Q^{\rm tot}_{\rm e,inj}$. These pairs upon acceleration obtain a power-law spectrum that extends from $\gamma \sim 1$ to $\gamma \sim \sigma_{\rm e}$ with a slope of $\sim -1$ in the limit of high magnetizations \cite{2001ApJ...562L..63Z,2008ApJ...682.1436L,2022PhRvL.128n5102S}. For simplicity, we call these pairs $N^{\rm trap}_{\rm X}$ although their spectrum is not only determined by acceleration at the X-points. The differential injection rate then reads

\begin{equation}
q^{\rm trap}_{\rm X}(\gamma<\gamma_{\rm inj})\equiv\frac{dN^{\rm trap}_{\rm X}}{d\gamma dt}=(1-\zeta)\frac{Q^{\rm tot}_{\rm e,inj}}{\ln(\sigma_{\rm e})}\gamma^{-1},
\label{eq:trapped_pairs_dis_approx_below_sigma_e}
\end{equation}
where the normalization factor $1/\ln(\sigma_{\rm e})$ is found by requiring that 
\begin{equation} 
\int_{1}^{\sigma_{\rm e}} d\gamma \ q^{\rm trap}_{\rm X}(\gamma) + \int_{\sigma_{\rm e}}^{\gamma_{\rm rad}}  d\gamma  \ q^{\rm trap}_{\rm fr}(\gamma) = (1-\zeta) Q_{\rm e, tot}^{\rm inj}.
\end{equation}

The steady-state distribution of X-trapped particles can be computed as \cite{Inoue_Takahara}
  \begin{equation}
        N_{\rm X}^{\rm trap}(\gamma) = e^{- \frac{\gamma^{\rm syn}_{\rm cool}}{\gamma}} \frac{\gamma^{\rm syn}_{\rm cool} t_{\rm adv}}{\gamma^2} \int_{\gamma}^{\infty} d\gamma \, q^{\rm trap}_{\rm X}(\gamma) \,  e^{\frac {\gamma^{\rm syn}_{\rm cool}}{\gamma}}, \gamma < \gamma_{\rm inj}
        \label{eq:X_trapped_steady-state}
    \end{equation}

\subsection{Numerical approach}\label{sub_sec:Numerical_appr}
For the numerical calculation of the non-thermal radiation from the free and trapped pairs in the reconnection region we utilize the code {\tt LeHaMoC}\footnote{\url{https://github.com/mariapetro/LeHaMoC/}} \citep{2024A&A...683A.225S}: a versatile time-dependent lepto-hadronic modeling code that includes synchrotron emission and absorption, inverse Compton scattering (on thermal and non-thermal photons), $\gamma \gamma$ pair production, photopair (Bethe-Heitler pair production) and photopion production processes, as well as proton-proton inelastic collisions. For this work, we will utilize the leptonic module of the code. 

The code has been designed to describe the time evolution of the particle distribution within a spatially homogeneous spherical source. Particles that are injected into the source are assumed to instantaneously fill the whole volume, while all particles escape from the source on the same timescale (i.e., their radial position from the center of the source is not considered when computing their geometric escape). Moreover, the emissivities of all processes are angle-averaged assuming isotropic distributions for charged particles and photons. 

\begin{figure*}%
    \centering
\includegraphics[width=0.9\textwidth]{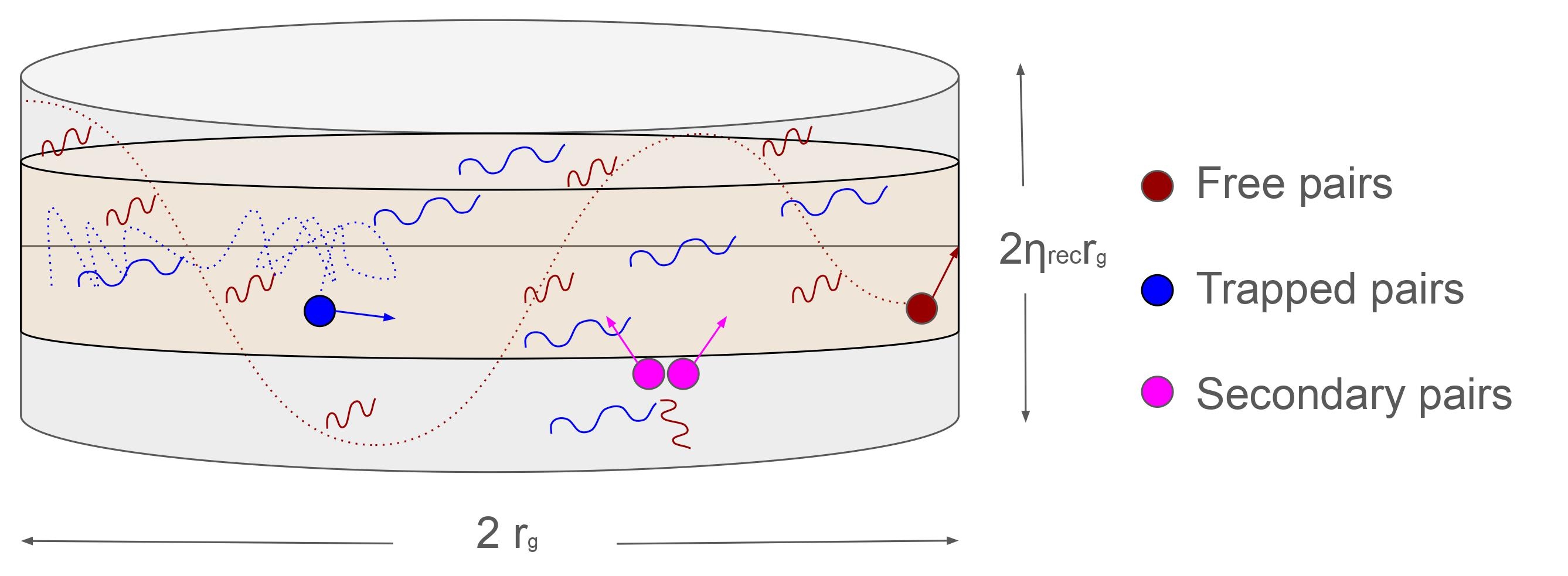}      
\caption{Schematic illustration of our model showing the cylindrical current sheet and the distinct pair populations. Photons emitted by a specific pair population are indicated by wavy lines of the same color. Blue and red dotted lines visualize the trajectory of a trapped and free particle, respectively.}
\label{fig:sketch}%
\end{figure*}

Therefore we do not account for the different spatial distributions of the free and trapped populations and their emitted photons in our numerical model (all are assumed to occupy the same region since we can not distinguish the upstream and downstream regions). Nonetheless, their spatial differences are encoded in the fact that free particles undergo active acceleration, while trapped particles do not. Moreover, with the adopted numerical framework we cannot directly study potential anisotropic radiative signatures \cite{2023ApJ...959..122C}. To ensure some equivalency between the physical current sheet and the numerical emitting region (in regards to density-dependent interaction rates), we consider a sphere of effective radius  $ R_{\rm eff} \simeq 0.5 R \eta_{\rm rec, -1}^{1/3}$ that has the same volume as the cylindrical current sheet of radius $R$ and height $2 \eta_{\rm rec} R$ (motivated by the width of the largest plasmoids in the layer, see Fig.~\ref{fig:sketch}).
Finally, to capture the realistic escape of trapped pairs, which happens along the current sheet on the plasma advection timescale, we set their escape timescale equal to $R/c$. Secondary pairs are mainly produced from the interaction of synchrotron photons from the free population with the other photons in the system. We assume that they are injected within the system's volume, do not undergo the free acceleration phase, and are considered part of the trapped population. Since they are trapped in the midplane, we assume they cool via synchrotron, inverse Compton losses, and escape the system in $R/c$ with $R$ being the half-length of the current sheet.


\section{Application to M87*}\label{sec:m87}
In this section, we apply our numerical model for radiation from magnetospheric current sheets to the SMBH M87*. We first compute the non-thermal radiation from the free and trapped populations, accounting also for inverse Compton scattering and $\gamma \gamma$ pair production, for different values of the upstream pair magnetization. We then set limits on the latter parameter using multi-wavelength observations of M87*, and discuss the implications of our findings for the pair enrichment of the upstream plasma and UHE proton acceleration (Sec.~\ref{sec:spectra}). 

\begin{table}
\centering
\caption{Model parameters and values used for the application to M87*.}
\begin{threeparttable}
\begin{tabular}{l c c}
\hline 
Parameter & Symbol & Value [Units]\\ \hline
Distance & $d_{\rm M87}$ & 16.4 [Mpc]\tnote{*}\\
Black hole mass     & $M$ & $6.5 \cdot 10^9$ [$M_{\odot}$]\\
Dimensionless mass accretion rate  & $\dot{m}$ &  $10^{-6}-10^{-5}$\tnote{**}\\
Current sheet half-length & $R$ &  1 [$r_{\rm g}$] \\
Effective radius & $R_{\rm eff}$ &  0.5 [$r_{\rm g}$] \\
Reconnection rate & $\eta_{\rm rec}$ &  0.06  \\ 
Upstream magnetic field & $B_0$ &  $70-221$ [G]\tnote{\textdagger}  \\
Pair magnetization & $\sigma_{\rm e}$ & $10-10^6$ \tnote{\textdaggerdbl}  \\
Fraction of free particle injection & $\zeta$ & 0.06 \\
Multiple of $r_{\rm g}$ & $\mathcal{R}$ & 1 \\
Matter-to-luminosity conversion factor & $\eta_{\rm c}$ & 0.1 \\
Spin of the SMBH & $\alpha_{\rm s}$ & 1 \\

\hline 
\end{tabular}
\begin{tablenotes} 
\item[*] Adopted from  \cite{distance_m87}. 
\item[**] Motivated by modeling of polarimetric observations  \cite{Akiyama_2021}.
\item[\textdagger] Estimated using Eq.~\ref{eq:magn_field} for the range of $\dot{m}$ values listed here. 
\item[\textdaggerdbl] Motivated by the regime of interest $\sigma_e < \gamma_{\rm rad}\sim 10^6$ (see Eq.~\ref{eq:gamma_rad}). Such low values imply that the magnetospheric region, where current sheets form, has pair densities much greater than the Goldreich-Julian value.  This could materialize if some plasma is channeled from the disk to the layer in the dynamically evolving environment of MAD, or/and if pairs are generated by $\gamma \gamma$ absorption as described in \cite{2011ApJ...735....9M}. 
\end{tablenotes}
\end{threeparttable}
\label{tab:param}
\end{table}

\subsection{Photon emission from free and trapped pairs}\label{sec:spectra}
We compute the steady-state\footnote{We note that the trapped pair population reaches a steady state only asymptotically. In practice, after $\sim 10R_{\rm eff}/c$ the densities reach 99.9954 percent of their asymptotic values.} distributions of free and trapped pairs, as well as the emitted photon spectra using the code {\tt LeHaMoC}. All results are obtained at $T=15 R_{\rm eff}/c$, assuming that this is the relevant lifetime of the current sheet. The parameters are listed in Table~\ref{tab:param}.

\begin{figure*}%
    \centering
\includegraphics[width=0.49\textwidth]{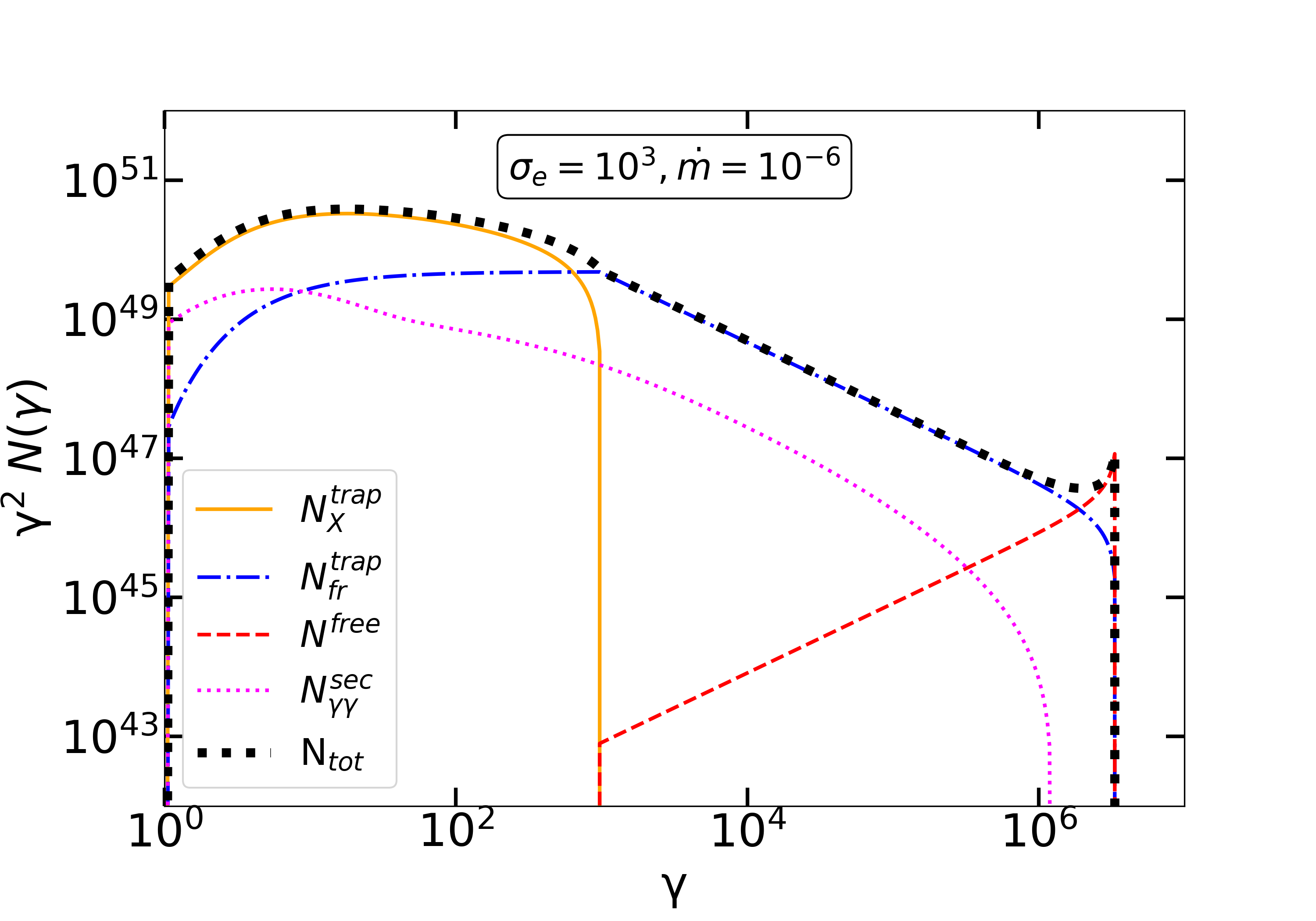} 
\includegraphics[width=0.49\textwidth]{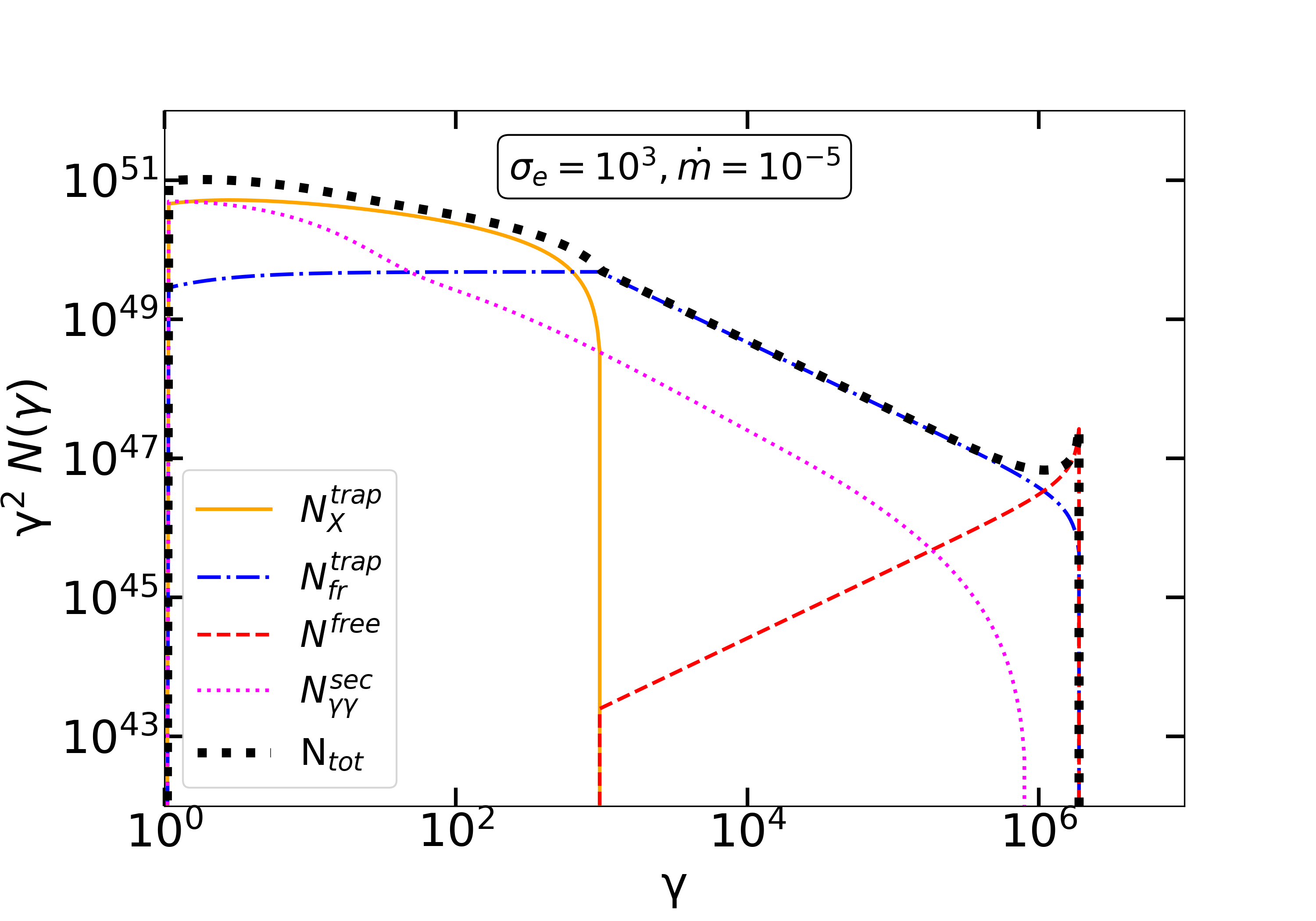} 
\includegraphics[width=0.49\textwidth]{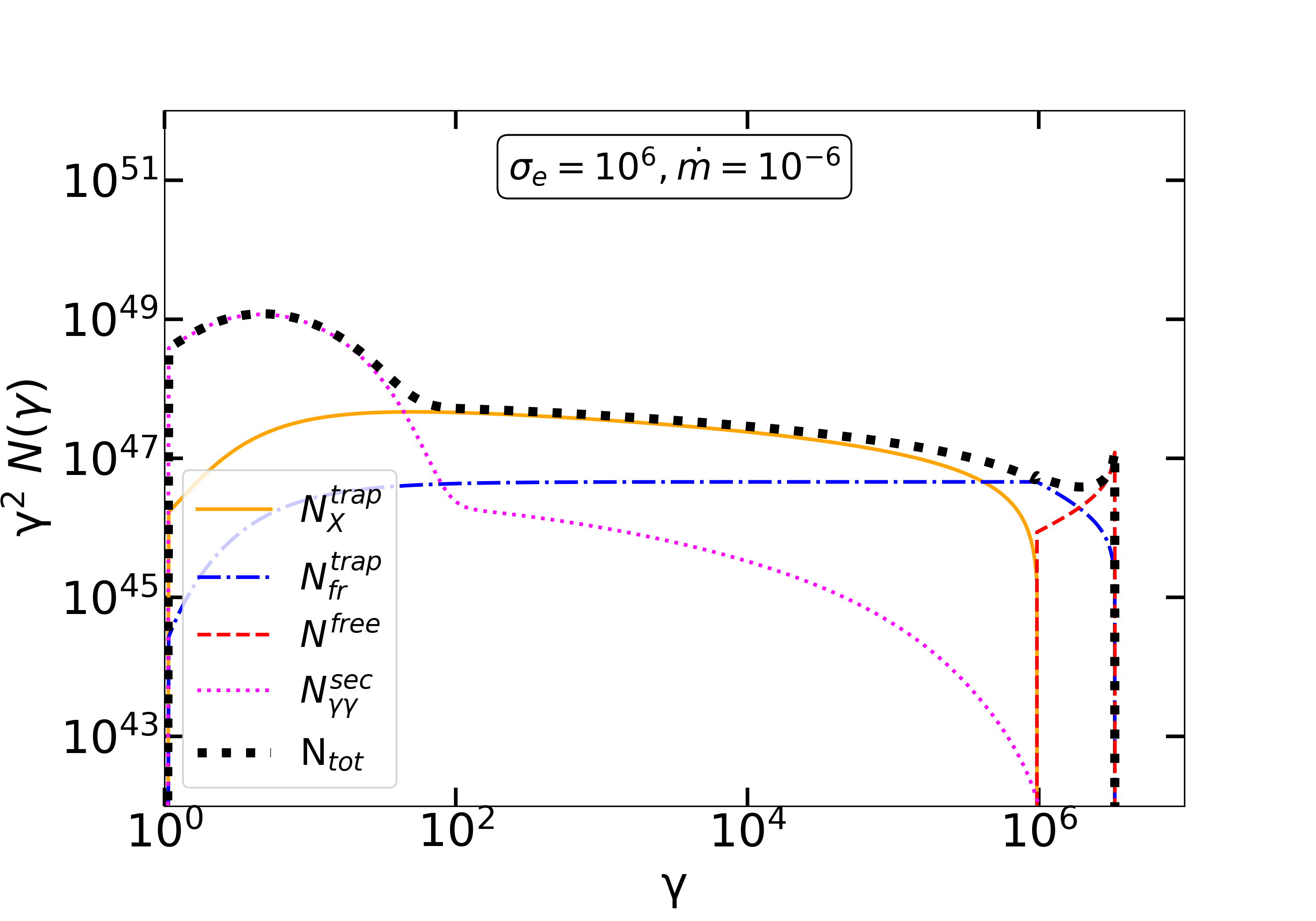} 
\includegraphics[width=0.49\textwidth]{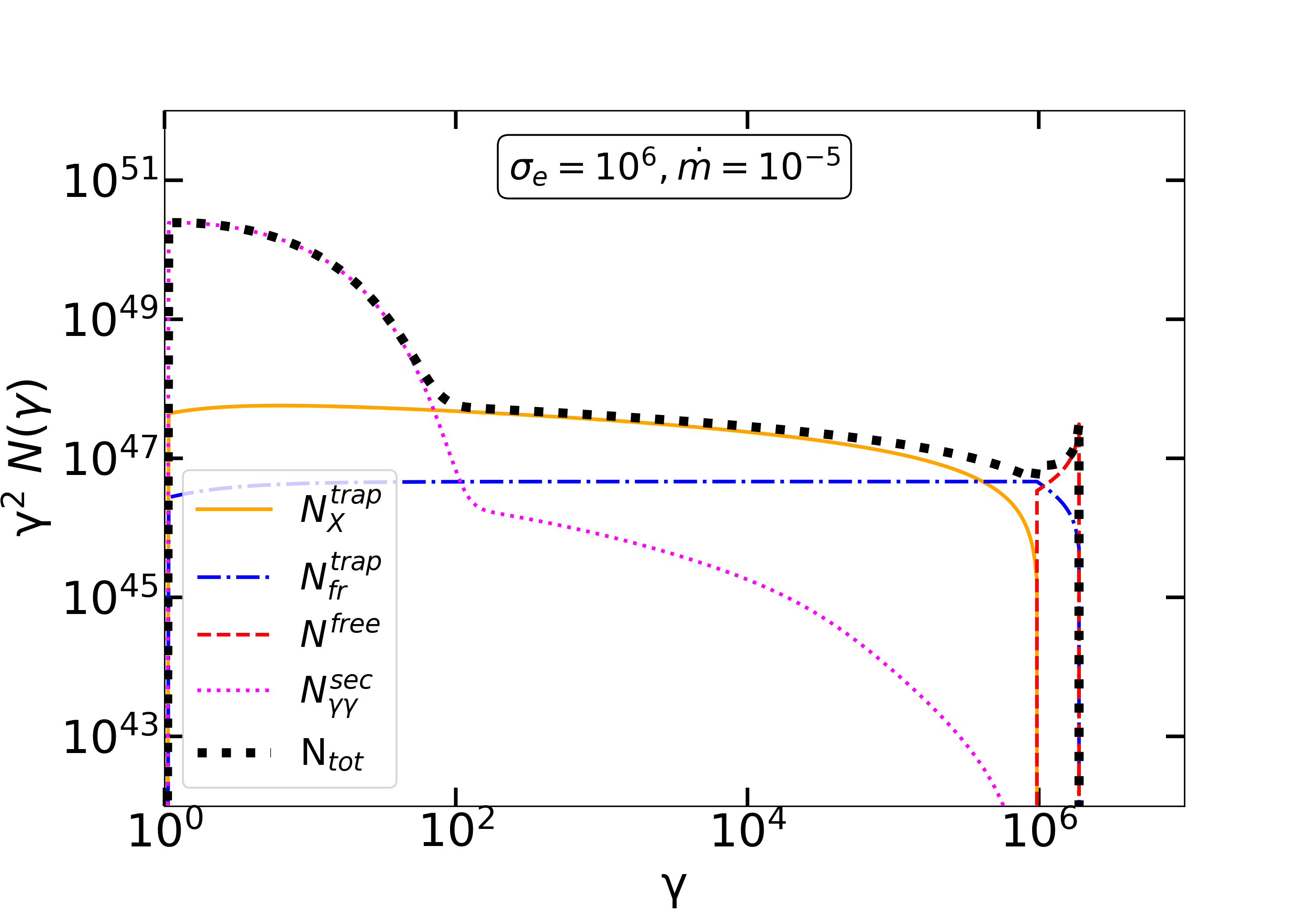} 
    \caption{Decomposition of the steady-state pair distribution for $\sigma_{\rm e}=10^3$ (top) and  $\sigma_{\rm e}=10^{6}$ (bottom) for different accretion rates, $\dot{m}=10^{-6}$ (left) and $\dot{m}=10^{-5}$ (right). The attenuation of the 10~MeV free synchrotron photons produces secondary pairs with $\gamma \sim 10$, while more energetic pairs are injected by the attenuation of inverse Compton scattered photons.}  
    \label{fig:e_spec_decomp_mdot} 
\end{figure*}

\begin{figure*}%
    \centering
\includegraphics[width=0.48\textwidth]{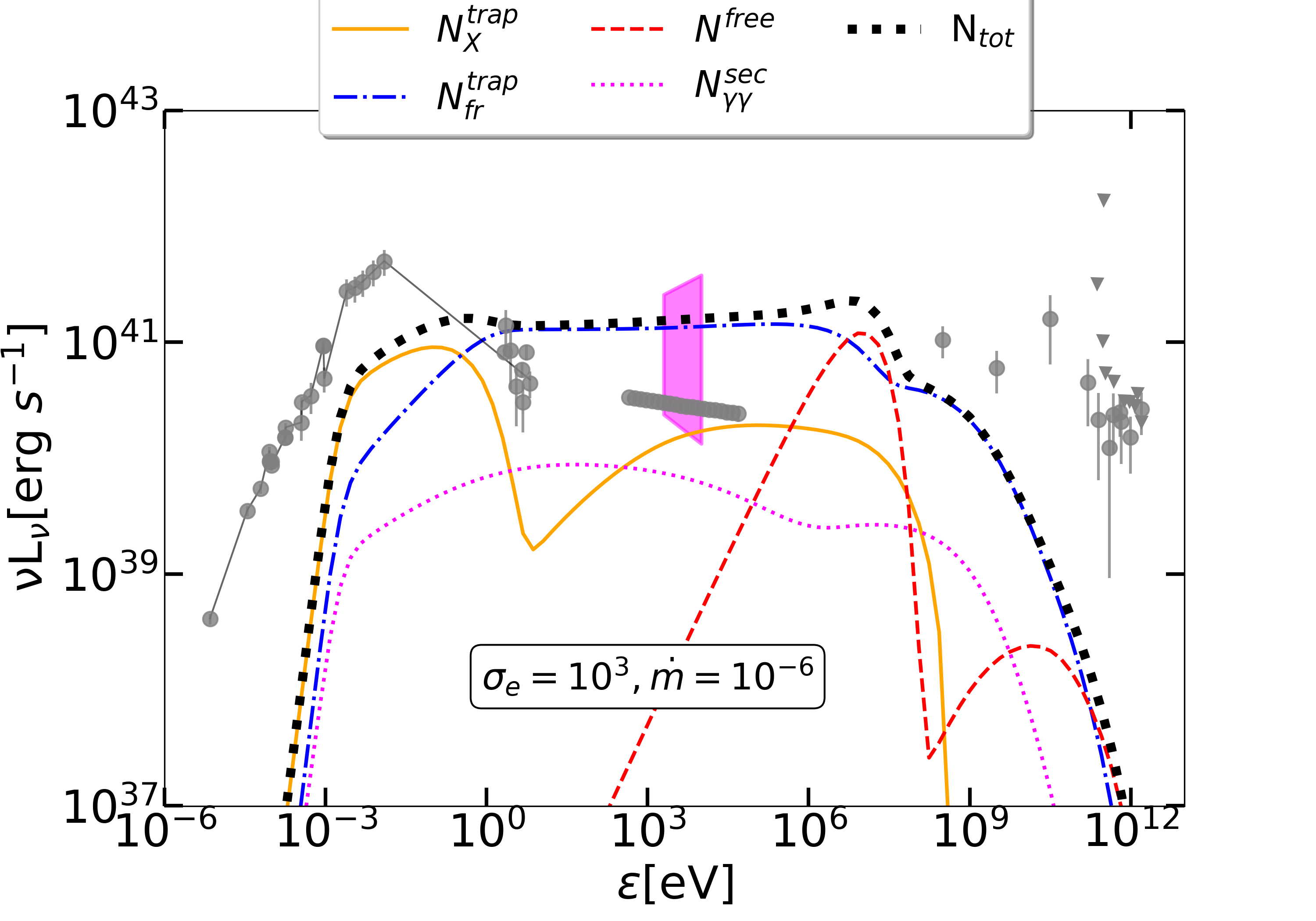} 
\includegraphics[width=0.48\textwidth]{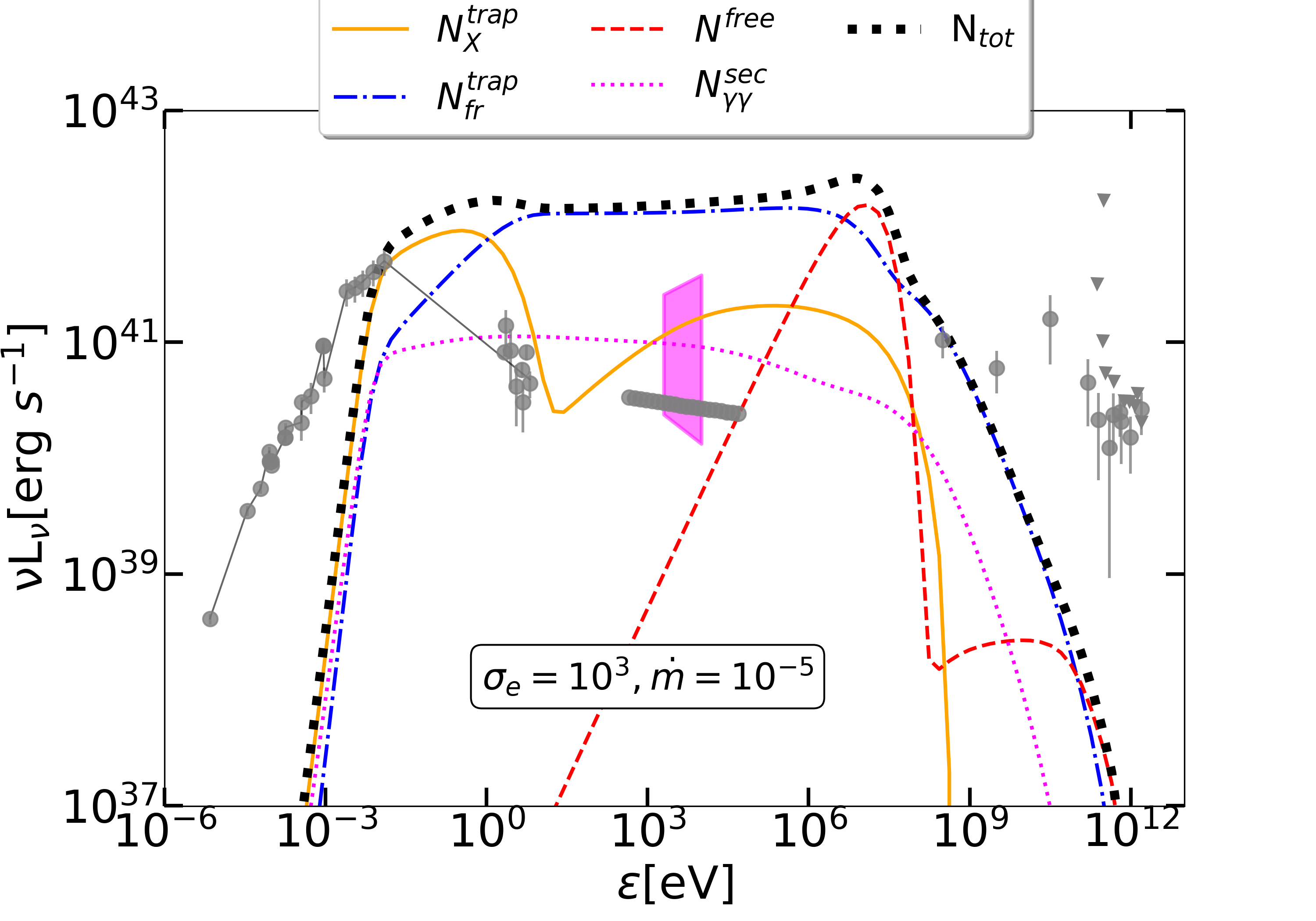} 
\includegraphics[width=0.48\textwidth]{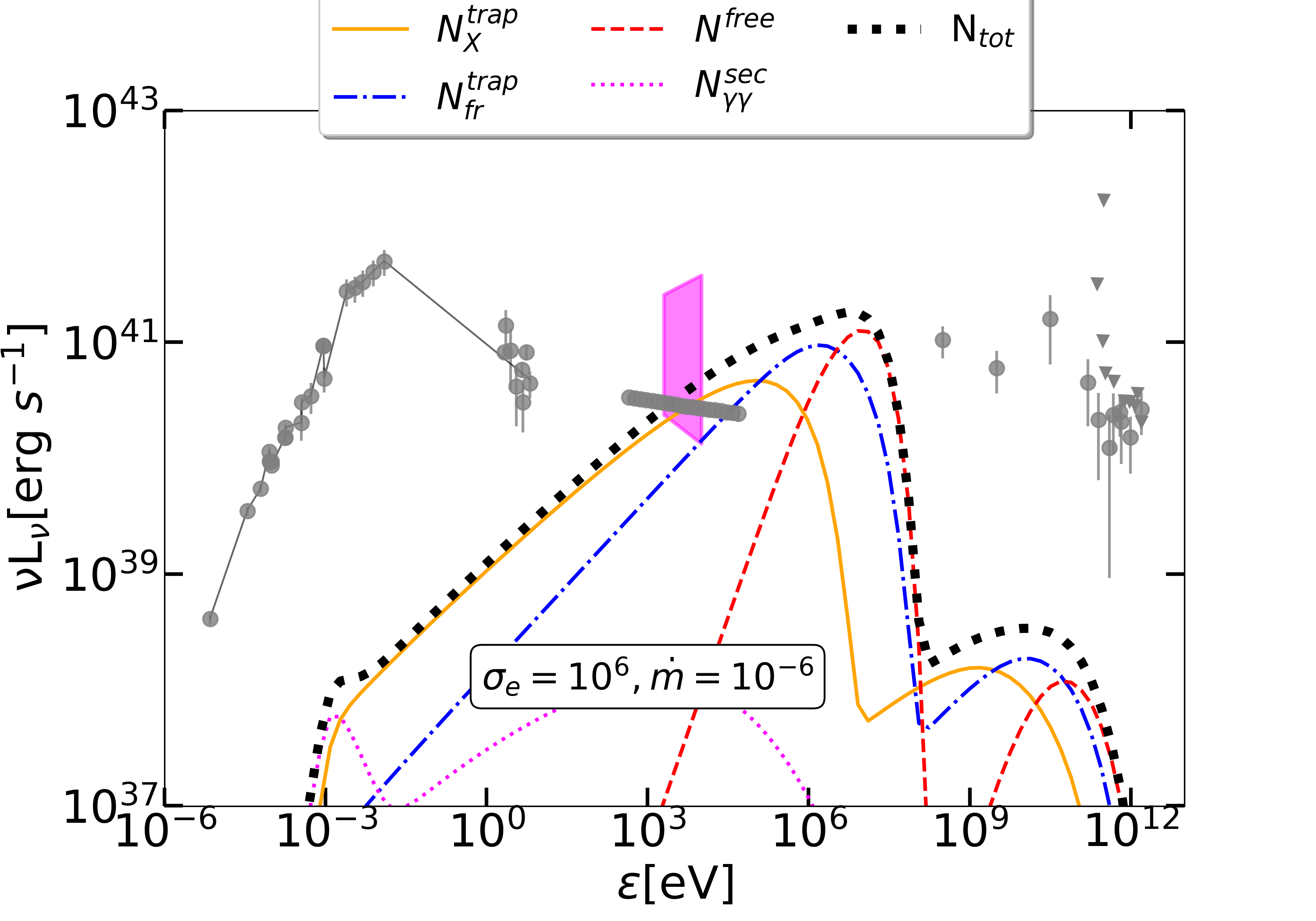} 
\includegraphics[width=0.48\textwidth]{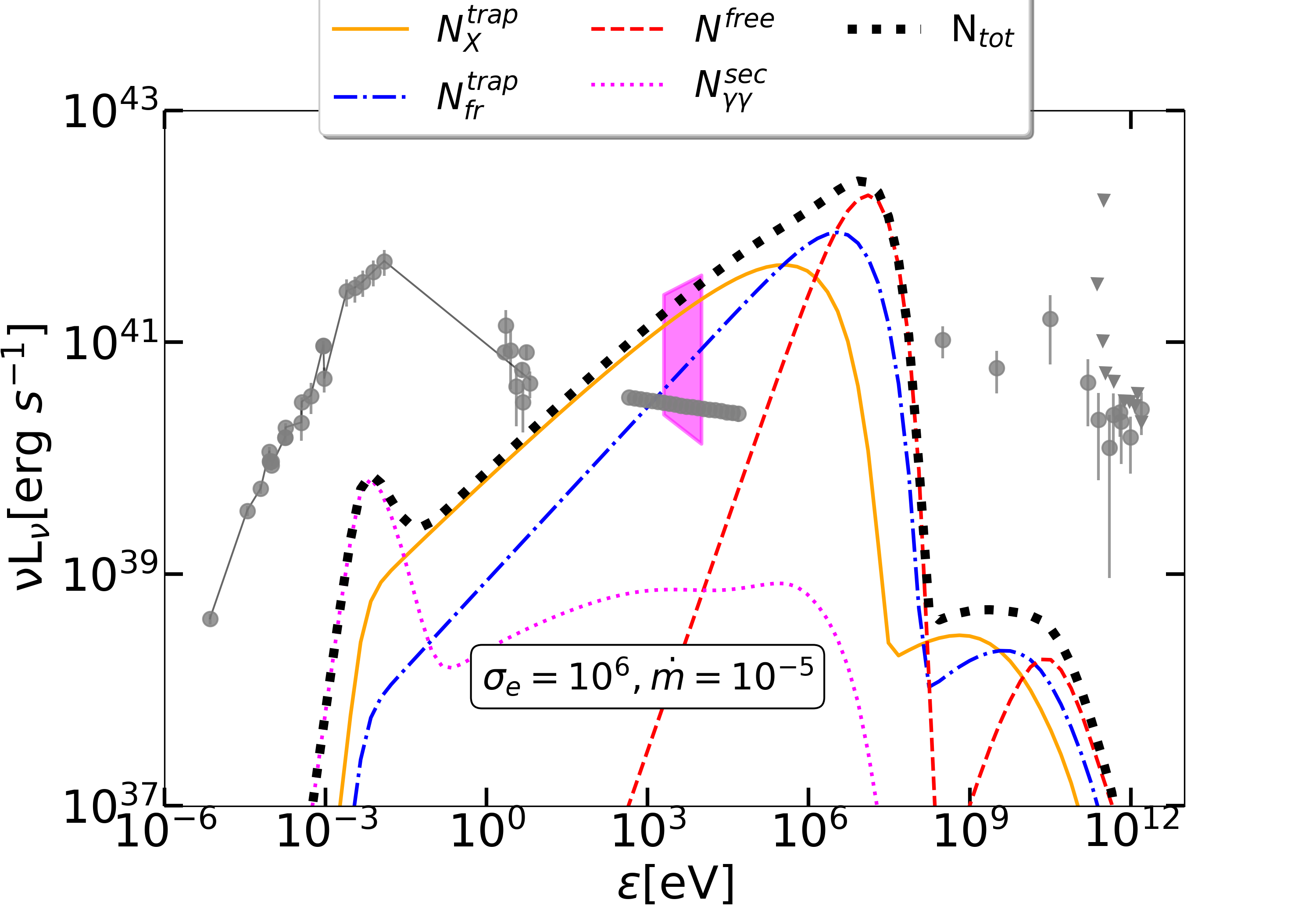} 
    \caption{Decomposition of the steady-state photon spectrum for $\sigma_{\rm e}=10^3$ (top) and  $\sigma_{\rm e}=10^{6}$ (bottom) for different accretion rates, $\dot{m}=10^{-6}$ (left) and $\dot{m}=10^{-5}$ (right). Grey points show archival flux measurements of M87* from Refs. \citep{refId0,Algaba_2021}, and      the magenta-shaded area consists of data from  Ref.~\citep{2023RAA....23f5018C}.}  
    \label{fig:ph_spec_decomp_mdot} 
\end{figure*}

In Fig.~\ref{fig:e_spec_decomp_mdot} we illustrate the decomposition of the pair spectrum under two distinct accretion rate scenarios ($\dot{m}=10^{-6}$ on the left and $\dot{m}=10^{-5}$ on the right), for two different values of the magnetization, $\sigma_{\rm e}=10^3$ (top) and $\sigma_{\rm e}=10^3$ (bottom). The red dashed line in all plots represents the population of free pairs $N^{\rm free}(\gamma)$ accelerated in the upstream region from $\gamma_{\rm inj}\approx \sigma_{\rm e}$ to $\gamma_{\rm rad}$ with a power law index $\sim -1$. The acceleration process leads to a notable spike in the distribution at $\gamma_{\rm rad}$, indicating the accumulation of pairs in this energy range. The pile-up at the radiation-limited Lorentz factor is expected to grow with the lifetime of a current sheet. The small variation in the position of $\gamma_{\rm rad}$ between cases with different $\dot{m}$ stems from the weak dependence of $\gamma_{\rm rad}$ on the accretion rate as outlined in Eq.~(\ref{eq:gamma_rad}). The blue dashed-dotted line and the yellow solid line depict the distributions of trapped pairs originating from two distinct channels, i.e. $N^{\rm trap}_{\rm fr}(\gamma)$ and $N^{\rm trap}_{\rm X}(\gamma)$ respectively. The total distribution of trapped pairs can be described by a broken power law, where the break happens at $\gamma\sim \sigma_{\rm e}$. Above $\sigma_{\rm e}$ we see the contribution of the $N^{\rm trap}_{\rm fr}$ population while below the break the contribution of $N^{\rm trap}_{\rm X}$ pairs dominates. This pair population is cooled down to $\gamma^{\rm syn}_{\rm cool}$ (see Eq.~\ref{eq:syn_cool}) forming a flat spectrum in $\gamma^2 N(\gamma)$ (see yellow line in all panels of Fig.~ \ref{fig:e_spec_decomp_mdot}). Finally, the magenta dotted line in both panels indicates the steady-state spectrum of secondary pairs produced via $\gamma \gamma$ pair production. The main channel for secondary pair production is the attenuation of synchrotron photons with energy $\epsilon_{\rm syn}\sim 10$~MeV, which are emitted by the free pairs with $\gamma \simeq \gamma_{\rm rad}$ (see red dashed lines in Fig.~\ref{fig:ph_spec_decomp_mdot}). The attenuation of the 10 MeV photons will lead to the production of pairs with $\gamma^{\rm sec}_{\rm inj} \approx \epsilon_{\rm syn}/(2 m_{\rm e} c^2) \approx 10$, where most of the energy is also injected; see bump at $\gamma\sim 10$ in the left panels of Fig.~\ref{fig:e_spec_decomp_mdot}. Secondary pairs are also injected with $\gamma\sim 10$ in the right column of Fig.~\ref{fig:e_spec_decomp_mdot} but their distribution is modified by cooling more strongly due to the larger magnetic field. The peak of the secondary pair energy spectrum scales almost quadratically with $\dot{m}$, because both the target photon luminosity and the 10 MeV photon luminosity depend linearly on $\dot{m}$, see Eqs.~(\ref{eq:nLn_syn}) and (\ref{eq:L_sec}). For lower $\sigma_{\rm e}$ values we expect the production of more low energy photons compared to cases with higher $\sigma_{\rm e}$ (see also Fig.~\ref{fig:ph_spec_decomp_mdot}). These photons are upscattered to higher energies, i.e., above 10 MeV up to TeV energies (see Fig.~\ref{fig:ph_spec_decomp_mdot}). The attenuation of these photons is responsible for the production of secondary pairs above $\gamma>10$ up to $\gamma \sim 10^6$. The energy of photons upscattered to TeV energies is larger for the cases with lower $\sigma_{\rm e}$ we find that the energy that goes into the secondary pairs from the attenuation of TeV photons is more significant for lower $\sigma_{\rm e}$ values.

Fig.~\ref{fig:ph_spec_decomp_mdot} presents a decomposition of the photon spectrum emitted by different pair populations for the same parameters used for Fig.~\ref{fig:e_spec_decomp_mdot}. We also show with grey markers multi-wavelength observations of M87* \citep{refId0,Algaba_2021}. The grey solid line in Fig.~\ref{fig:ph_spec_decomp_mdot} represents the photon field from the inner accretion flow of M87* as assumed in our numerical model (which is included as a source of soft photons for inverse Compton scattering). For
energies between $\sim 4 \cdot 10^{-3}$~eV and 4 eV where no data are available, we assume that the photon spectrum is a power law whose index is benchmarked by UV observations\footnote{All flux measurements, besides the EHT measurement at 230 GHz, which is spatially resolved, should be considered as upper limits on the flux in the reconnection region as they are likely produced at larger distances from the SMBH.}. The magenta-shaded region is constructed using X-ray observations taken over a 12-year-long period with Chandra from the core region of M87 \cite{2023RAA....23f5018C}. In general, the X-ray spectrum observed by Chandra is harder when brighter. The red dashed line represents the photon spectrum produced by free pairs $N^{\rm free}(\gamma)$ via synchrotron and inverse Compton scattering. The synchrotron emission peaks at 10~MeV, originating from pairs with $\gamma\simeq \gamma_{\rm rad}$ and is the primary source of photons in this energy range.  The blue dashed-dotted line and the yellow solid line illustrate the photon spectrum emitted by trapped pairs ($N^{\rm trap}_{\rm fr}$ and $N^{\rm trap}_{\rm X}$ respectively), including synchrotron and inverse Compton processes. Similar to the pair distributions, photons emitted by trapped pairs exhibit a broken power-law spectrum. Above $\epsilon_{\rm syn}(\sigma_{\rm e}) \approx 6~{\rm eV}~\sigma_{\rm e,3}^2 \dot{m}_{-5}^{1/2} M_9^{-1/2} \eta_{\rm c,-1}^{-1/2}f^{-2}(a_{\rm s})$, we observe the contribution to the total emission from $N^{\rm trap}_{\rm fr}$, where the spectrum is flat in $\nu L_{\nu}$ since these photons are produced by cooled pairs with power law index ~$\sim  -3$, while at lower energies the spectrum has both $N^{\rm trap}_{\rm fr}$ and $N^{\rm trap}_{\rm X}$ contributions. Synchrotron photons from free pairs are mainly absorbed by the photons originating from $N^{\rm trap}_{\rm fr}$ and serve as the source of secondary pairs $N^{\rm sec}_{\rm \gamma \gamma}$, as discussed in the previous paragraph. Because of the dependence of the secondary energy injection rate on $\dot{m}$ (see Eq.~\ref{eq:L_sec}), their synchrotron emission will be more luminous for higher accretion rates (compare magenta dotted lines in left and right panels of Fig.~\ref{fig:ph_spec_decomp_mdot}). The synchrotron spectra of secondary pairs are limited by synchrotron self-absorption at photon energies below $\sim 10^{-3}$~eV, and their contribution to the total luminosity is negligible compared to the emission from free-trapped pairs above $\epsilon_{\rm syn}(\sigma_{\rm e})$.
Synchrotron photons can be upscattered by the same pairs to higher energies\footnote{Each pair population ``sees'' all non-thermal photons (produced by all pair populations) as targets for inverse Compton scattering. We also take into account the disk photons, whose energy density is computed within a sphere of radius $R_{\rm soft}> R_{\rm eff}$  as $U_{\rm soft}=3 \, L_{\rm soft}/(4\pi R_{\rm soft}^2c)\simeq 0.04~\mathrm{erg \, cm^{-3}}$. Here, $L_{\rm soft} \approx \nu_{\rm EHT} L_{\nu_{\rm EHT}} \simeq 10^{41} $~erg s$^{-1}$, $\nu_{\rm EHT} =230$~GHz, and $R_{\rm soft} = 5 r_{\rm g}$.}. Because of the low energy density of the target photons compared to the energy density of the magnetic field the produced luminosity of the inverse Compton scattered photons is typically $\sim 1-2$ orders of magnitude lower than the synchrotron luminosity depending on $\sigma_{\rm e}$.  

\begin{figure*}%
    \centering
\includegraphics[width=0.49 \textwidth]{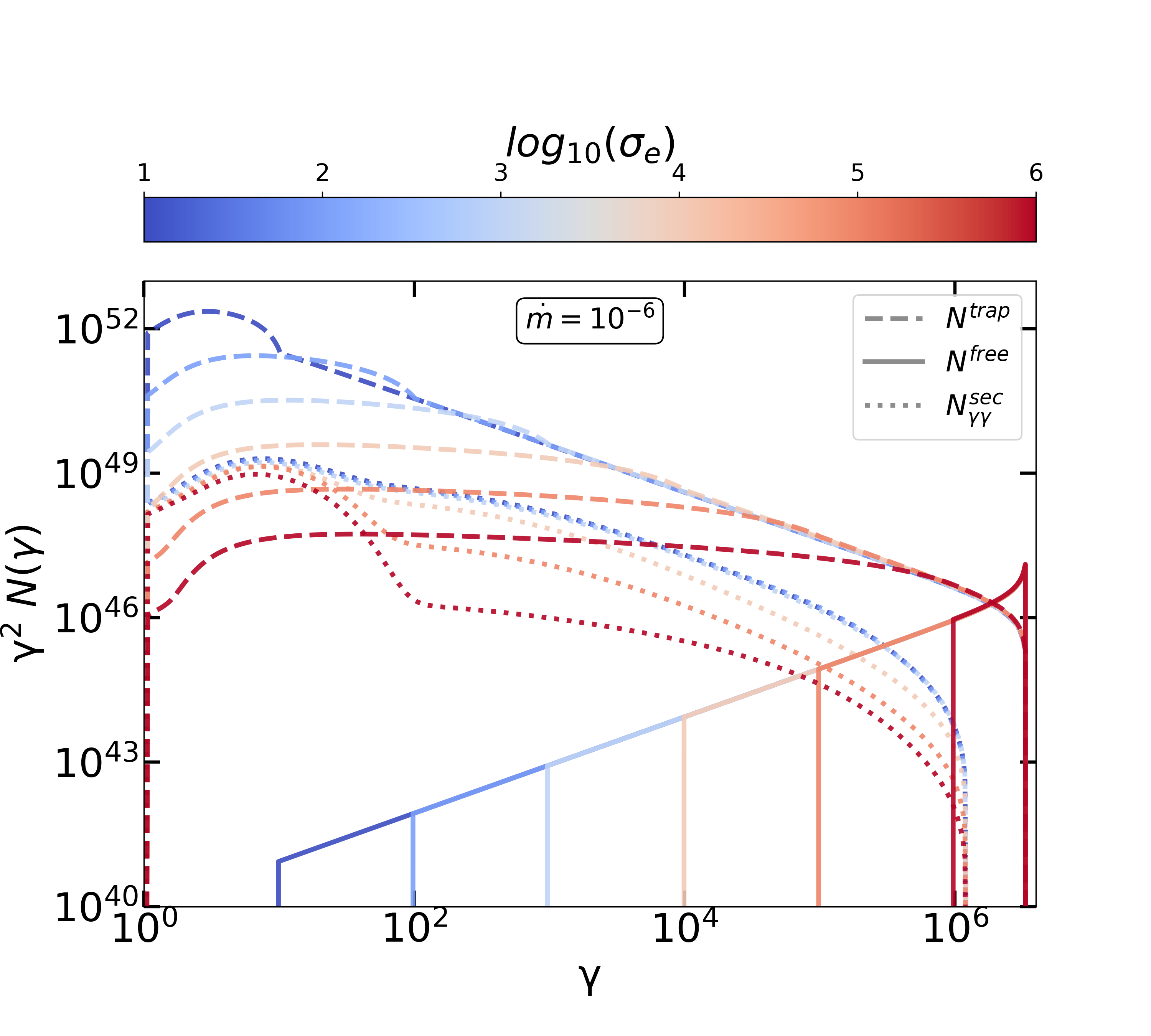}     
\includegraphics[width=0.49 \textwidth]{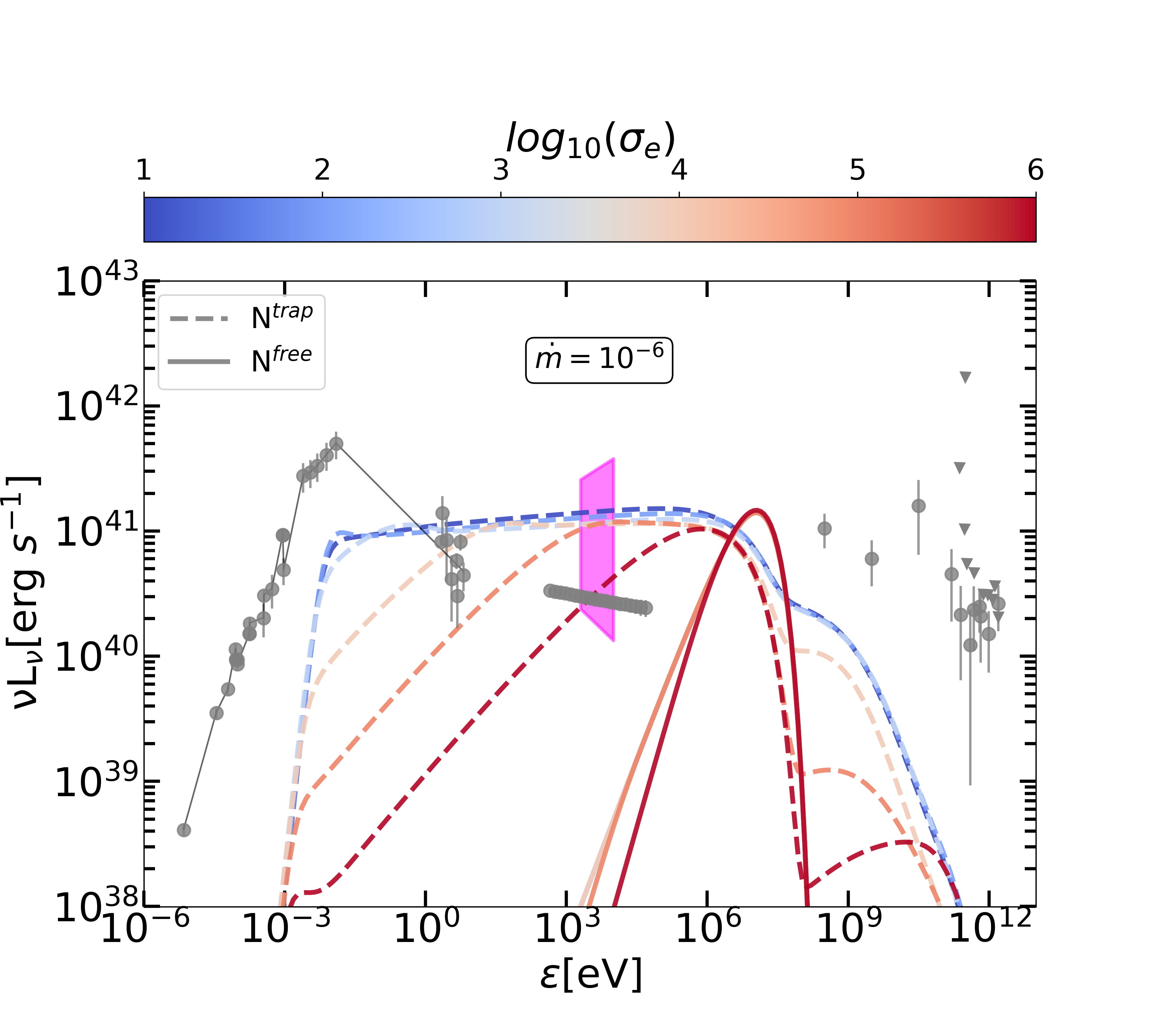}
\caption{Steady-state pair distribution (left panel) and non-thermal photon spectrum (right panel) computed for $\dot{m}=10^{-6}$ and different values of the upstream pair magnetization, as shown in the color bar.}
\label{fig:e_ph_spec_vs_sigma_free_vs_trapped}%
\end{figure*}

The left panel of Fig.~\ref{fig:e_ph_spec_vs_sigma_free_vs_trapped} depicts the overall distribution of free pairs (solid lines) and trapped pairs (dashed lines) at steady state obtained for different initial values of the pair magnetization ($\sigma_{\rm e}$), while maintaining a constant value of $\dot{m}=10^{-6}$. Therefore, the upstream magnetic field is fixed in all cases, and by changing $\sigma_{\rm e}$ we effectively change the upstream number density of pairs. At this stage,  we do not account for the contribution of secondary pairs into the definition of $\sigma_{\rm e}$. Dotted lines indicate secondary pairs generated by $\gamma \gamma$ absorption. The bump of the secondary pair spectrum at $\gamma \sim 10$ remains unaffected because these pairs originate by the attenuation of the 10~MeV photons whose luminosity and energy are independent of $\sigma_{\rm e}$. Nonetheless, the number of more energetic secondary pairs ($\gamma \gg 10$) increases with decreasing $\sigma_{\rm e}$ due to the increase in the inverse Compton scattered photon luminosity (see right panel). On the right panel, we demonstrate the resulting photon spectra, showing the free and trapped pair contributions separately. In summary, higher magnetization values lead to harder X-ray spectra (right panel) and higher number densities of secondary pairs relative to the initial pair density of the system (left panel).

\subsection{Constraints from observations}\label{sub_sec:con_f_obv}
Neglecting at the moment the contribution of secondary pairs in the total number density of plasma (i.e., $\sigma_{\rm e}$ is defined based on the primary pair density), we can use X-ray observations to constrain $\sigma_{\rm e}$ for a given $\dot{m}$. We remind that the non-thermal emission of the current sheet is transient, lasting about a day, assuming a lifetime of $T = 15 (R_{\rm eff}/c) \simeq 2.6$~days. A comparison to observations obtained on similar timescales is, therefore, the most meaningful.

\begin{figure}
\centering
\includegraphics[width=0.48\textwidth]{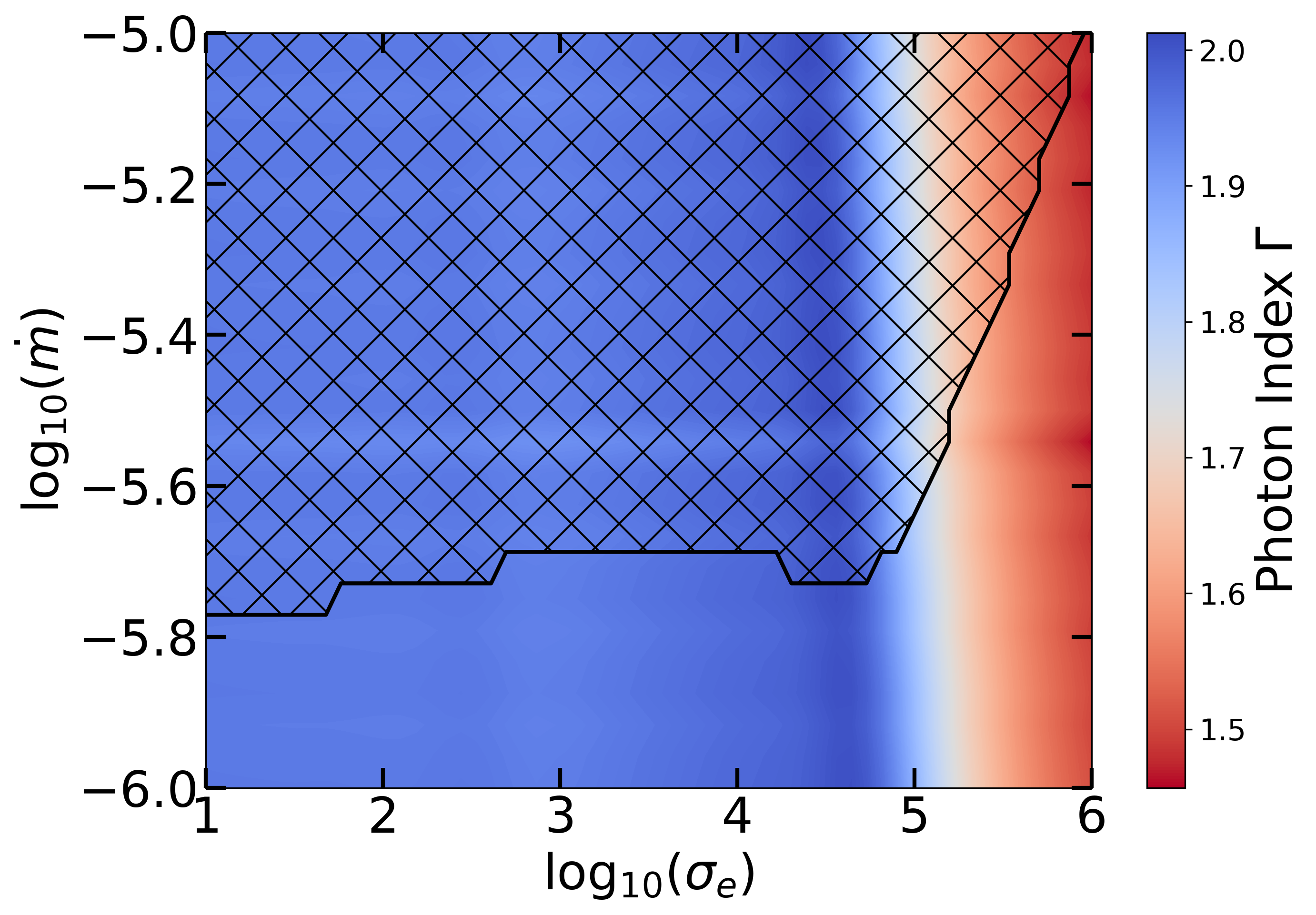} 
\includegraphics[width=0.48\textwidth]{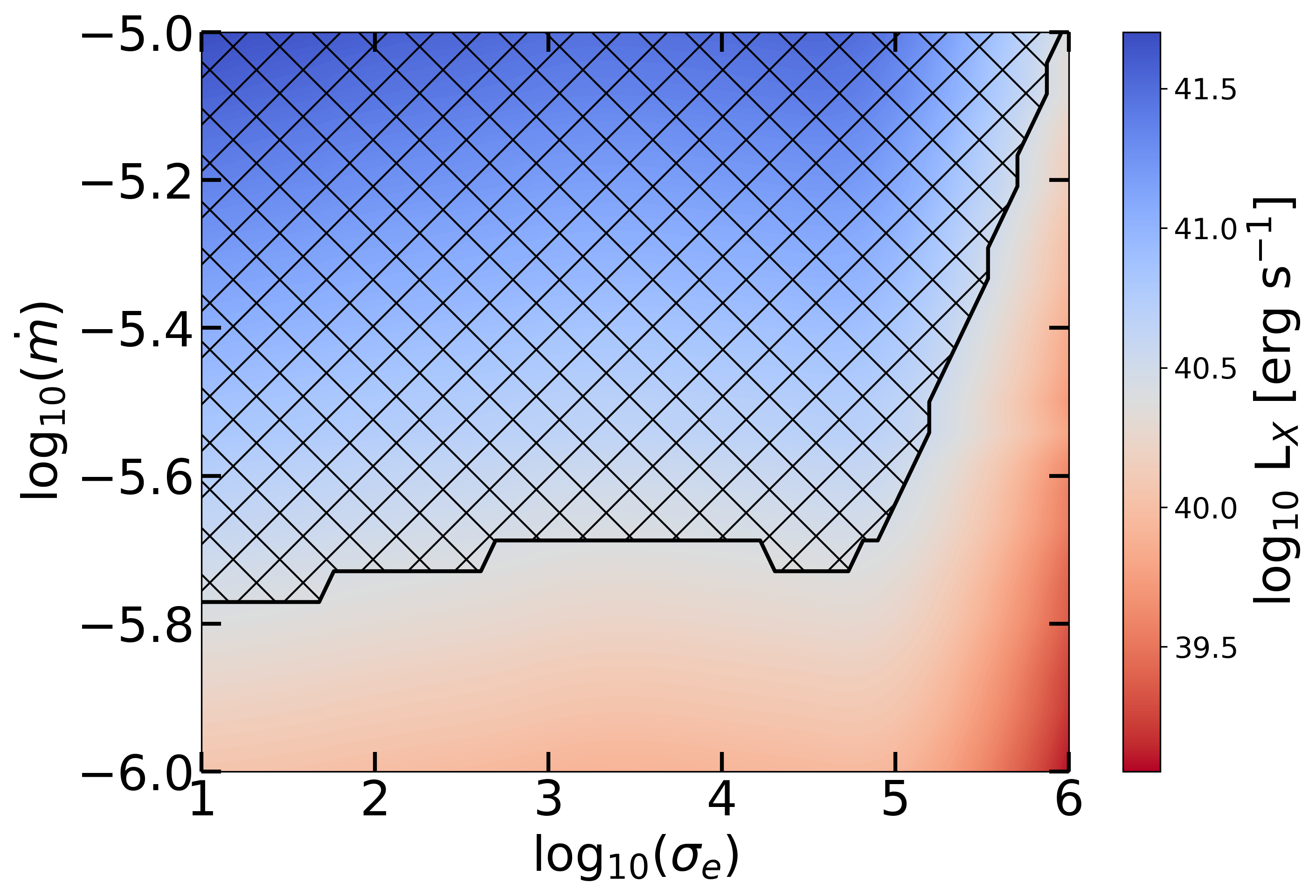} 
\caption{Left panel: Photon index (color bar) of non-thermal spectra in 0.2-10~keV energy range, obtained for different combinations of $\dot{m}$ and $\sigma_{\rm e}$. The area outside the mesh indicates combinations for which the predicted luminosity falls within the magenta-shaded area in Fig. \ref{fig:e_ph_spec_vs_sigma_free_vs_trapped}. Right panel: X-ray luminosity $L_{\rm X}$ (color bar) of non-thermal spectra in 0.2-10~keV energy range, obtained for different combinations of $\dot{m}$ and $\sigma_{\rm e}$. The area outside the mesh indicates combinations for which the predicted luminosity falls within the magenta-shaded area in Fig. \ref{fig:e_ph_spec_vs_sigma_free_vs_trapped}. 
}
\label{fig:mdot_sigma}
\end{figure}

By considering 4.7~ks X-ray observations from 31st, 2007 to March 28th, 2019 with Chandra/ACIS \cite{2023RAA....23f5018C} (see the magenta shaded area in Fig.  \ref{fig:e_ph_spec_vs_sigma_free_vs_trapped}) we can determine which combinations of $\dot{m}$ and $\sigma_{\rm e}$ result in spectra within the observed range. We adopt the range of accretion rate values provided by Ref.~\cite{Akiyama_2021}. In Fig.~\ref{fig:mdot_sigma} we present the photon index\footnote{The photon index $\Gamma$ is defined as the slope of the differential photon energy spectrum and is calculated by performing a linear fit to the quantity $\log_{10}(\nu F_{\nu}) \propto (-\Gamma+2) \log_{10}(\nu)$ as derived by the model in the desired energy range}. map of the synchrotron pair spectrum in the 0.2-10 keV energy range (see color bar) for various combinations of $\dot{m}$ and $\sigma_{\rm e}$. For $\sigma_{\rm e} \lesssim 10^4$ we find soft X-ray spectra, as the emission is dominated by the cooled part of the trapped pair population. Instead, for higher magnetizations the spectra become harder, reaching the asymptotic value of $3/2$. The non-hatched region indicates solutions that fall inside the observed range of X-ray spectra (see the magenta-shaded region in Fig.~\ref{fig:e_ph_spec_vs_sigma_free_vs_trapped}). Similarly, we show the 0.2-10 keV X-ray luminosity map as predicted by our model for a combination of $\sigma_{\rm e}$ and $\dot{m}$ values. Inspection of both panels shows that only a small portion of the considered parameter space is consistent with the spread of X-ray observations from the core region of M87. For $\sigma_{\rm e} \lesssim 10^5$, only the lowest accretion rates are viable producing a flat spectrum ($\Gamma \simeq 2$), while accretion rates as high as $10^{-5}$ are plausible for $\sigma_{\rm e} \sim 10^6$.

\subsection{UHE proton acceleration}\label{sub_sec:UHECR}
Ultra-high energy acceleration in the vicinity of SMBHs in AGN has been widely discussed (for a recent review, see \cite{Rieger_2022}). Here, we examine if acceleration by the reconnection-driven electric field in magnetospheric current sheets can push protons to energies above $10^{18}$~eV.

To determine whether acceleration at magnetospheric current sheets in M87* can push protons to ultra-high energies, one must compare the acceleration timescale with the various loss timescales relevant for relativistic protons. Knowledge of the target photon density and spectrum is needed for photomeson and photopair (Bethe-Heitler) production processes. 

We consider first the disk photons. Apart from the EHT measurement flux at 230 GHz, whose origin is spatially resolved on a radial scale $\lesssim 7 r_{\rm g}$, all other fluxes are likely produced at larger distances in the accretion flow. We consider the next two limiting cases: (i) a high photon-density scenario, where all the observed flux emanates from the same region as the 230 GHz flux, and (ii) a low photon-density scenario, where just the 230~GHz photons from the inner accretion flow act as targets for proton cooling. In addition to the photons from the accretion flow, we take into account the non-thermal photons emitted by the free and trapped pairs in the reconnection region (as described in  section~\ref{sec:spectra}).

The characteristic proton timescales (for their definitions, see Appendix~\ref{app:times}) are plotted in Fig.~\ref{fig:M_87_timescales_p} against the proton Lorentz factor for typical parameter values (see Table~\ref{tab:param}). The yellow- and red-colored bands indicate respectively the range of synchrotron loss and acceleration timescales obtained by using the lowest  (solid lines) and highest (dashed lines) values of the mass accretion rate, as inferred from the modeling of polarimetric observations \citep{Akiyama_2021}. The blue- and magenta-shaded regions demonstrate the range of values for the photopion and Bethe-Heitler energy loss timescales, respectively, considering only the disk photons. Dashed-dotted lines mark the longest timescales obtained when only the 230~GHz photons are considered as targets (lowest density scenario), while solid blue and magenta lines indicate the shortest timescales obtained assuming that radio-to-UV fluxes emanate from the same region as the current sheet (highest density scenario). In Fig.~\ref{fig:M_87_timescales_p} we also show the timescales of the photopion and Bethe-Heitler loss timescales, when accounting for both, the high photon-energy density scenario and the non-thermal emission of pairs (dotted blue and magenta lines respectively) for $\sigma_{\rm e} = 10^4$ and $\dot{m}=10^{-6}$. These choices of parameters combine the observational constraints and the regulation of the magnetization coming from the pair creation, as will be discussed in the following section.

\begin{figure}
\centering
\includegraphics[width=0.8\textwidth]{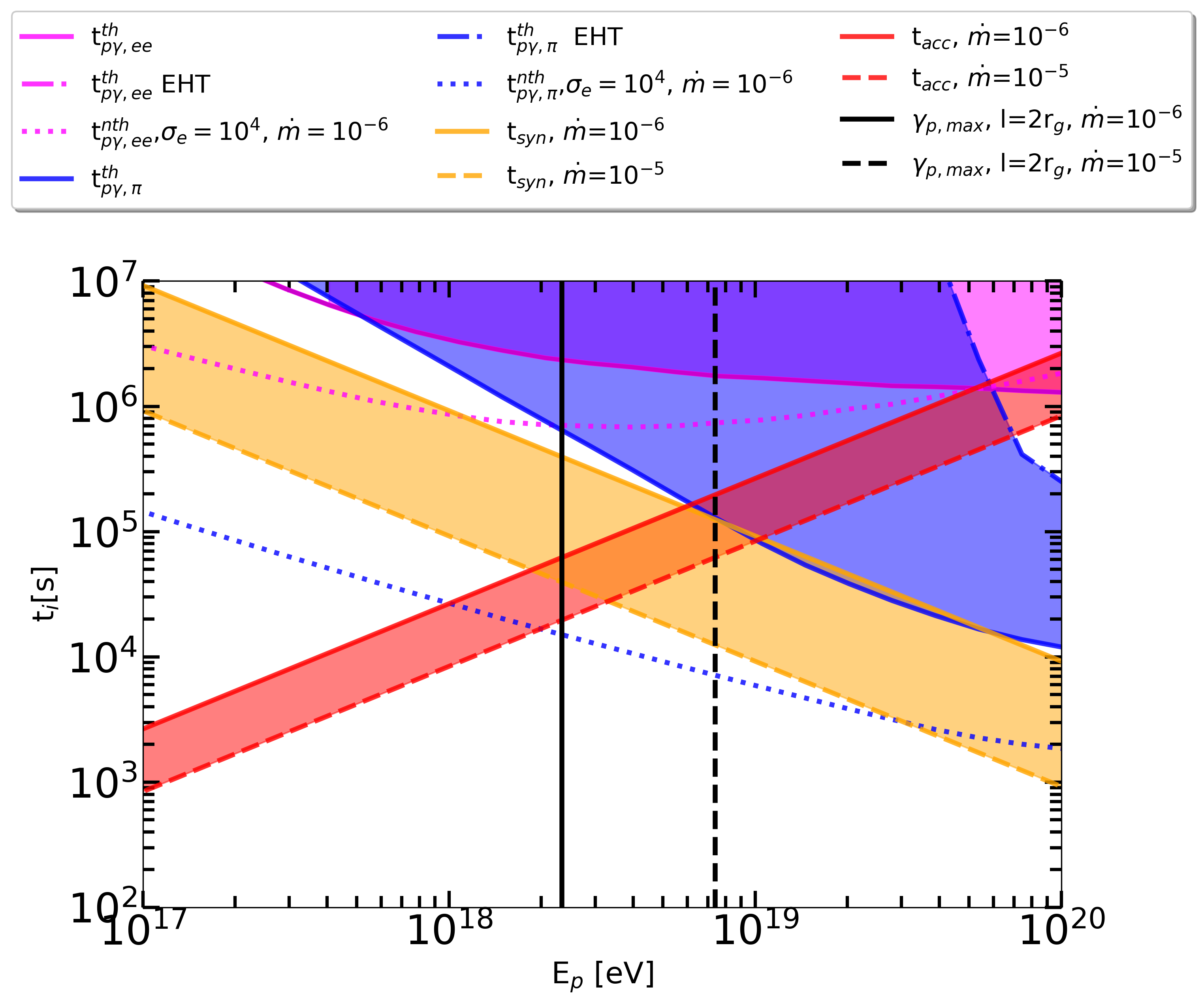}
\caption{Characteristic timescales for relativistic protons accelerated in a magnetospheric current sheet of M87*. The acceleration and synchrotron timescales are represented respectively by red- and orange-shaded regions, which are derived assuming a range for the mass accretion rate, $\dot{m} \in (1-10)\times 10^{-6}$. Solid blue and magenta lines indicate the photopion and Bethe-Heitler loss timescales, respectively, in the high-density scenario, where all observed flux is assumed to originate from the same region as the current sheet. Dashed blue and magenta lines indicate the photopion and Bethe-Heitler loss timescales, respectively, when accounting for the non-thermal emission of pairs for $\sigma_{\rm e}=10^4$ and $\dot{m}=10^{-6}$. The dotted lines indicate the timescales in the low-density scenario, where only the EHT flux at 230~GHz is considered. The vertical solid and dashed lines mark the maximum Lorentz factor given by Eq.~(\ref{eq:gamma_p_max}) for two extreme values of the accretion rate. }
\label{fig:M_87_timescales_p}
\end{figure} 

For protons with $E_{\rm p} \lesssim 1$~EeV, the acceleration timescale (red band) is shorter than all energy loss timescales. Bethe-Heitler losses are always negligible for the range of Lorentz factors shown in the figure. For $\sigma_{\rm e}=10^{4}$ photopion losses on the non-thermal photons from pairs are dominating the proton energy losses, limiting their acceleration to energies $E_{\rm p}\lesssim 1$~EeV (crossing of the dotted blue line and red band). The radiation-limited proton energy, in this case, is lower than the maximum energy that can be achieved in a current sheet of length $l=2R$ and electric field $E_{\rm rec} = 0.1 \eta_{\rm rec,-1} B_0$,
\begin{equation}
E^{\rm (acc)}_{\rm p, \max} \simeq e \eta_{\rm rec} B_0 l\simeq 4~{\rm EeV} \, \frac{\eta_{\rm rec,-1}(\dot{m}_{-5}M_9)^{1/2}\mathcal{R}_0\eta_{\rm c, -1}^{-1/2}}{f^2(a_{\rm s})}.
\label{eq:gamma_p_max}
\end{equation}
Higher magnetizations, e.g. $\sigma_{\rm e}>10^4$, would lead to lower number densities of target photons (see Fig.~\ref{fig:e_ph_spec_vs_sigma_free_vs_trapped}), shifting the blue dotted line upwards, and making synchrotron losses the relevant limiting mechanism of proton acceleration.

\begin{figure}
\centering
\includegraphics[width=0.7\textwidth]{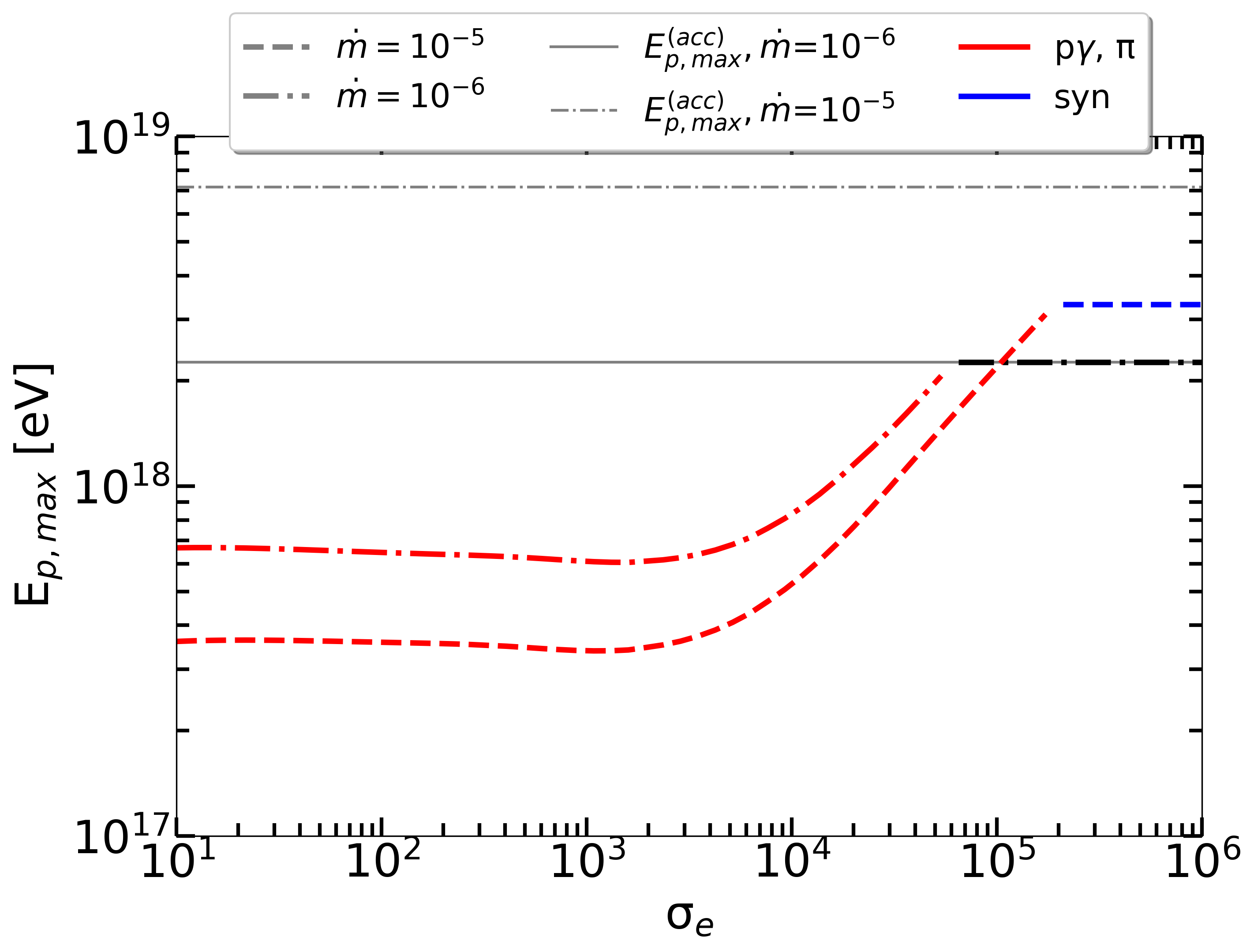}
\caption{Maximum energy reached by protons in a magnetospheric current sheet of length $2 r_{\rm g}$ in M87* for two values of $\dot{m}$ and different magnetization values $\sigma_{\rm e}$.}
\label{fig:E_p_max}
\end{figure}

In Fig.~\ref{fig:E_p_max} we show the maximum energy achieved by protons, namely $\min[E^{\rm (acc)}_{\rm p,\max}, E^{\rm (rad)}_{\rm p,\max}]$ where $E^{\rm (rad)}_{\rm p,\max}$ denotes the radiation limited energy of protons due of cooling, across various magnetization values $\sigma_{\rm e}$ and for two distinct mass accretion rates $\dot{m}$. It is evident that $E^{\rm (rad)}_{\rm p,\max}$ is contingent upon both parameters, while $E^{\rm (acc)}_{\rm p,\max}$ is independent of $\sigma_{\rm e}$ (see grey solid and dashed-dotted line in Fig.~\ref{fig:E_p_max}). For magnetization $\sigma_{\rm e}  \lesssim 10^3$ we observe that the maximum energy achieved by protons, which is limited by photopion energy losses, is almost constant because the target photon field does not differ much for these values of magnetization (see e.g. Fig.~\ref{fig:e_ph_spec_vs_sigma_free_vs_trapped}). The dependence of $E^{\rm (rad)}_{\rm p,\max}$ on $\dot{m}$ for $\sigma_{\rm e} \lesssim 10^3$ is driven by the fact that the trapped pairs and their synchrotron target photon density are higher for larger accretion rates as higher $\dot{m}$ values correspond to a higher magnetic field (refer to Eq. \ref{eq:magn_field}). Therefore the acceleration is limited to lower energy values for larger $\dot{m}$. For $10^3 \lesssim \sigma_{\rm e} \lesssim 10^5$, proton acceleration is also limited by photopion energy losses. The rising trend of $E_{\rm p,\max}$ in the range $10^3 \lesssim \sigma_{\rm e} \lesssim 10^5$ is attributed to the change of the minimum injection energy of trapped pairs $\sim \sigma_{\rm e} m _{\rm e} c^2$, which is reflected upon the distribution of target photons produced by these pairs (see the spectral break as $\sigma_{\rm e}$ increases in Fig~\ref{fig:e_ph_spec_vs_sigma_free_vs_trapped}). As $\sigma_{\rm e}$  increases further, for both values of $\dot{m}$, we find a constant $E_{\rm p,\max}$. For the low accretion rate, protons gain the maximum energy available into the system as given by Eq.~\ref{eq:gamma_p_max} (see black dashed-dotted line in Fig.~\ref{fig:E_p_max}). Conversely, for higher accretion rates and thus stronger magnetic fields, synchrotron losses become the limiting factor in proton acceleration (see blue dashed line in Fig.~\ref{fig:E_p_max}).

\subsection{Pair enrichment}\label{sub_sec:pair_enr}
In Fig.~\ref{fig:e_ph_spec_vs_sigma_free_vs_trapped} we showed that, for fixed $\dot{m}$, the number of secondary pairs injected at $\gamma \lesssim 10$ becomes progressively larger than the number of trapped pairs for higher initial values of $\sigma_{\rm e}$. Therefore, secondary pairs may play a crucial role in the self-regulation of plasma magnetization (see also \cite{2022ApJ...937L..34K,2023ApJ...944..173C}). By employing Eqs. (\ref{eq:nLn_syn}), (\ref{eq:epsilon_l}), and (\ref{eq:gamma_sec}) and assuming that the target photon field for the attenuation of 10 MeV photons\footnote{For the number density of the target photon field we assume that is given from $n_{\rm l}=3L_{\rm syn}^{\rm pk}/(4\pi R^2c\epsilon_{\rm l}$) where in the calculation we assumed that $R=\mathcal{R}r_{\rm g}$.} is constant for all magnetizations we can show that the ratio between the secondary pair density ($n_{\rm \gamma \gamma}^{\rm sec}$) and the number density $n_{\rm e^{\pm}}$ corresponding to the initial value of $\sigma_{\rm e}$ (see Eq.~\ref{eq:sigma}) is given by
\begin{equation}
\frac{n^{\rm sec}_{\rm \gamma \gamma}}{n_{\rm e^{\pm}}}\simeq 0.1\ \sigma_{\rm e,3}\dot{m}_{-5}\mathcal{R}_0\eta_{\rm c,-1}\bigg(\frac{\zeta_{-1.2} \eta_{\rm rec,-1}}{f^2(a_{\rm s})}\bigg)^2.
\label{eq:ratio_eq}
\end{equation}
The above equation implies that larger current sheets and higher accretion rates can increase the density of secondary pairs compared to the initial pair density (in the upstream).

\begin{figure*}%
    \centering
\includegraphics[width=0.7\textwidth]{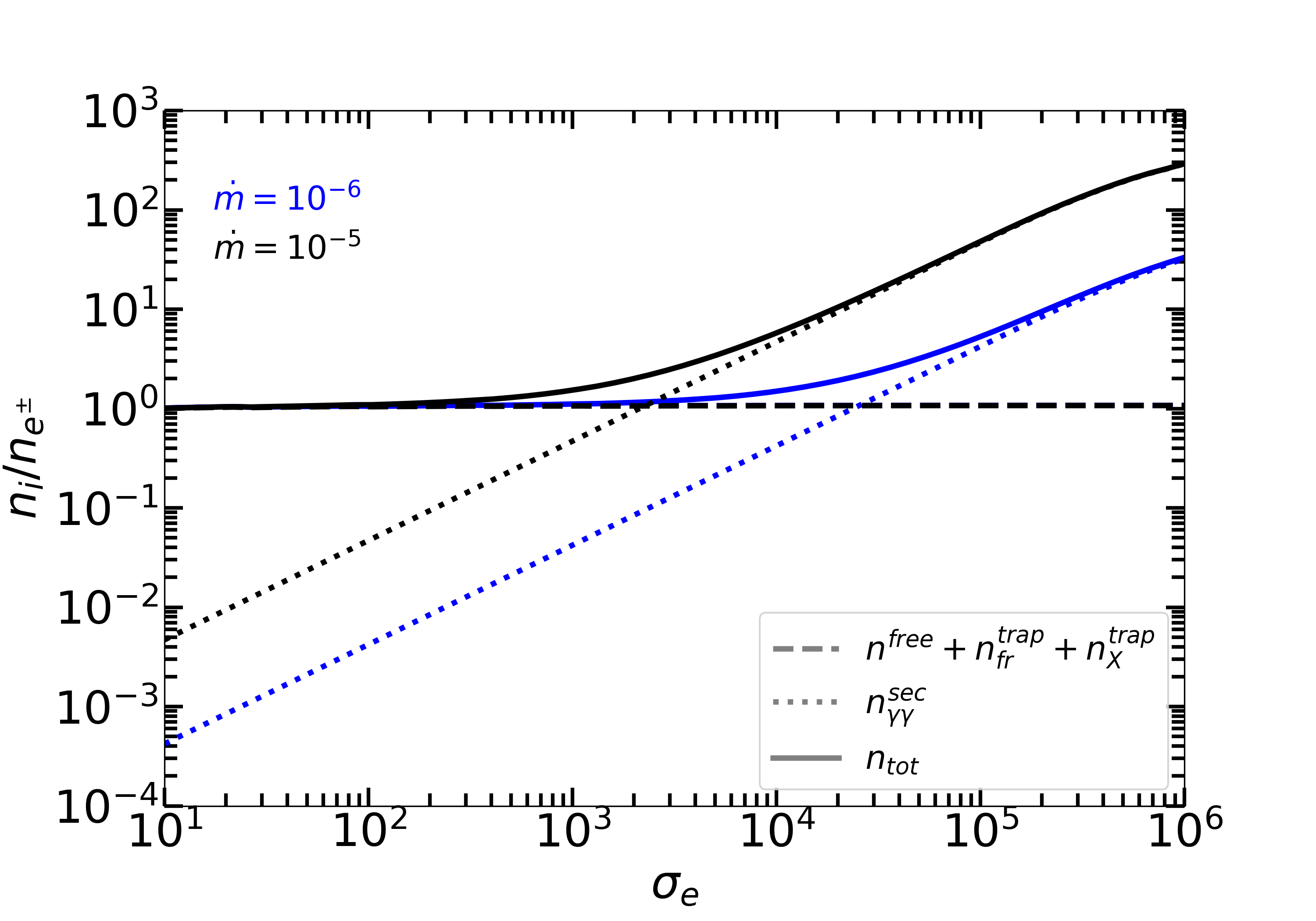} 
    \caption{Number density ratio between the total number density of primary and secondary pairs at steady state $n_{\rm i}$ and the initial pair density $n_{\rm e^{\pm}}$ as inferred by the adopted plasma magnetization $\sigma_{\rm e}$ for two different accretion rates, $\dot{m}=10^{-6}$ (blue lines) and $\dot{m}=10^{-5}$ (black lines). For high values of $\sigma_{\rm e}$ pair creation leads to $n_{\rm tot} \gg n_{\rm e^\pm}$, implying a dynamic impact on plasma magnetization.} 
    \label{fig:ratio_density_vs_sigma_e} 
\end{figure*}

In Fig. \ref{fig:ratio_density_vs_sigma_e} we present the ratio of the number density of primary ($n^{\rm free}+n^{\rm trap}_{\rm fr}+n^{\rm trap}_{\rm X}$) and secondary pairs ($n_{\rm \gamma \gamma}^{\rm sec}$), as measured at steady state (i.e., after 10 light crossing times), to the density $n_{\rm e^{\pm}}$ inferred by the chosen value of $\sigma_{\rm e}$. In both cases of $\dot{m}$, we note that the ratio equals unity when the assumed magnetization in the system is relatively low, signifying a scenario where secondary pairs are not going to affect the plasma composition. For higher values of the assumed $\sigma_{\rm e}$ (and a fixed magnetic field),  the initial pair number density in the system is lower, i.e., $n_{\rm e^{\pm}}\propto \sigma_{\rm e}^{-1}$, whereas the density of secondary pairs is independent of $\sigma_{\rm e}$ (see also Eq.~\ref{eq:L_sec}). Consequently, we anticipate the ratio to progressively rise linearly with $\sigma_{\rm e}$, and exceed unity. Beyond this point, the contribution of secondary pairs to the plasma density is not negligible, and we expect $\sigma_{\rm e}$ to dynamically evolve if the layer is sufficiently long-lived. The steady-state approach we adopted so far suggests that the system will eventually reach equilibrium at an effective magnetization $\sigma_{\rm e}^* \sim 3\cdot 10^4$ ($3\cdot 10^3$) for $\dot{m}=10^{-6}$ ($10^{-5}$). Nevertheless, a time-dependent approach where $\sigma_{\rm e}$ is evolving with time is needed to describe the effects of pair creation better. We refer the reader to Appendix \ref{app:sigma} for more details.

\section{Summary \& Discussion}\label{sec:discuss} 
We have introduced a framework to compute the non-thermal leptonic emission from magnetospheric current sheets inspired by recent results of 3D PIC simulations of relativistic magnetic reconnection in pair plasmas \cite{2021ApJ...922..261Z, 2023ApJ...956L..36Z} and GRMHD simulations of accreting SMBHs \cite{2022ApJ...924L..32R}. Our model relies on the formation of current sheets in the magnetospheric region of the SMBH. So far, GRMHD simulations showing the formation of these structures have been performed in the so-called MAD accretion regime. The particle distribution in the reconnection region consists of two distinct populations that are coined  ``free'' and ``trapped''. Free pairs undergo fast acceleration (on a timescale comparable to the particle gyration timescale) while meandering in the upstream region between the two sides of the current sheet. They leave the region of free acceleration on a timescale similar to their acceleration time, when they get captured by plasmoids in the current sheet, contributing to the trapped population. Additionally, pairs that went through the injection phase of acceleration without ever experiencing a free phase of acceleration are considered trapped. Therefore, trapped pairs are pre-accelerated and susceptible to radiative losses, while they can leave the system on the plasma advection timescale.

Utilizing the numerical code {\tt LeHaMoC} we computed the photon spectrum generated by these pair populations and quantified the enrichment of pairs in the upstream due to pair creation across a wide range of magnetization values ($\sigma_{\rm e} \in (10,10^6)$) and sub-Eddington mass accretion rates relevant to M87* ($\dot{m} \in (10^{-6},10^{-5})$). We showed that pair creation due to the attenuation of $\sim$ 10~MeV photons emitted by the free pairs can be substantial, thus effectively reducing $\sigma_{\rm e}$ from its initially high values (e.g. $>10^4$). In other words, if the initial plasma magnetization is $\gg 10^3$, then it will progressively reduce to $\sigma_{\rm e}^* \sim 300$ or $\sim 3\cdot10^3$ for $\dot{m}=10^{-5}$ or $10^{-6}$, respectively. While the system reaches this new equilibrium after hundreds of layer light crossing times \footnote{These lifetimes for the current sheet are not physically relevant, our analysis shows that within the first 10 light crossing times, we have the largest reduction of $\sigma_{\rm e}$.}, the reduction of magnetization within the first ten light crossing times can be significant (see e.g. Fig.~\ref{fig:sigma_evol} in Appendix \ref{app:sigma}). The enrichment of plasma due to $\gamma \gamma$ pair creation becomes more important for higher accretion rates and larger current sheets, as shown in Eq.~(\ref{eq:ratio_eq}).

We also numerically explored the non-thermal radiation arising from the free and trapped pair populations in a magnetospheric current sheet for parameters relevant to M87* and different initial magnetization values, accounting for synchrotron emission and absorption, inverse Compton scattering of the disk and synchrotron photons, and photon-photon pair production. For $\sigma_{\rm e} \lesssim 10^4$ the emission from the trapped pair population overshoots the optical/UV and X-ray flux measurements, unless $\dot{m} \sim 10^{-6}$ (or even lower). The emission from the free population of particles is, however, independent of $\sigma_{\rm e}$ and peaks at $\sim 10$~MeV energies, as shown in Fig.~\ref{fig:e_ph_spec_vs_sigma_free_vs_trapped}. A transient magnetospheric current sheet may therefore power a MeV photon flare with a hard spectrum and day-long duration (comparable to the lifetime of a current sheet with a length of a few gravitational radii). The luminosity of the flare depends on the reconnection rate, which is well constrained from PIC simulations, the fraction of particles participating in the free phase of acceleration, which can be benchmarked with PIC simulations, and on source parameters, like the length of the current sheet, the accretion rate and black hole mass and spin (see Eq.~\ref{eq:nLn_syn}). Note that the MeV flare luminosity does not depend on the unknown plasma magnetization in the magnetospheric region. Therefore, a MeV $\gamma$-ray monitoring instrument, like AMEGO-X \cite{2022JATIS...8d4003C}, would be ideal for testing the particle acceleration scenario in magnetospheric current sheets.

Our analysis also revealed that non-thermal radiation from pairs may limit the acceleration of protons in current sheets due to photopion losses. The proton radiation-limited energy depends on $\sigma_{\rm e}$ and $\dot{m}$ through the low-energy ($< 0.1$~eV for $E_{\rm p}\simeq10^{18}$~eV) photon number density when photopion losses are dominant, but also through the magnetic field whenever synchrotron losses dominate. The radiation-limited proton energy generally increases with increasing $\sigma_{\rm e}$ till it reaches the maximum energy possible by the reconnection electric field along the whole length of the current sheet. The latter can be as high as 20 EeV for the largest magnetospheric current sheets (length of $10 \ r_g$) in M87*. The length of the current sheet does not impact the target photon energy density. Therefore, the photomeson loss timescale will be the same for different sizes of the current sheet. Even though the available potential energy for proton acceleration is larger in larger current sheets,  their acceleration is limited by synchrotron losses. By utilizing Eq.~(\ref{eq:gmax-syn}) and Eq.~(\ref{eq:gamma_p_max}) we can show that $E_{\rm p, max}^{(\rm syn)}/E^{\rm (acc)}_{\rm p, \max} \simeq \mathcal{R}_0^{-1}M_9^{-1/4}\dot{m}_{-5}^{-3/4}\eta_{\rm c,-1}^{1/4}\eta_{\rm rec,-1}^{-1/2}f(a_{\rm s})^3 $ where $E_{\rm p, max}^{(\rm syn)}$ is the synchrotron radiation-limited proton energy. Therefore for larger current sheets and $\sigma_{\rm e}>10^5$ the acceleration will be limited by the synchrotron losses to $3$~EeV $M_9^{1/4} \eta_{\rm c,-1}^{-1/4} \dot{m}_{-5}^{-1/4} \eta_{\rm rec, -1}^{1/2} f(a_{\rm s})$ energies.

Sgr~A* has been discussed as a PeV particle accelerator \cite{2016Natur.531..476H}. We therefore discuss the implications of our model for particle acceleration in magnetospheric current sheets for Sgr~A*. In contrast to M87*, Sgr~A* possesses a lower mass $M_{\rm Sgr~A^*}\simeq 4\cdot 10^{6}M_{\odot}$ \cite{SraA_EHT}, and accretes at an even lower rate, i.e., $\dot{m} \in (10^{-8},10^{-7})$  (assuming $\eta_{\rm c}=0.1$) \cite{SraA_EHT}. Assuming that accretion onto the black hole happens in the MAD regime, the magnetic field strength in the magnetospheric region of Sgr~A*, as determined by Eq.~(\ref{eq:magn_field}), is comparable with the one in M87*. Given that the length of the current sheets is a multiple of $r_{\rm g}$, the electric potential drop along a magnetospheric current sheet in Sgr~A* will be lower than in M87*. The upper limit on the energy achieved by protons in Sgr~A* is found, using Eq.~(\ref{eq:gamma_p_max}), to be

\begin{equation}
E^{\rm (Sgr~A^*)}_{\rm p, \max} \simeq e \eta_{\rm rec} B_0 l\simeq 8~{\rm PeV} \, \frac{\eta_{\rm rec,-1}(\dot{m}_{-8}M_{\rm Sgr~A^*})^{1/2}\mathcal{R}_0\eta_{\rm c, -1}^{-1/2}}{f^2(a_{\rm s})}.
\label{eq:gamma_p_max_SgrA}
\end{equation}
This suggests that radiation losses will not restrict proton acceleration, unlike in M87*, enabling protons to tap the system's maximum energy potential. Therefore, magnetospheric current sheets in Sgr~A* can push protons to maximum energies of a few PeV.  

In the analyses carried out in Sections \ref{sec:Th_fr} and \ref{sec:m87}, we incorporated the influence of the supermassive black hole's spin, denoted as $a_{\rm s}$, across all evaluated physical properties. Assuming that current sheets may also form for slowly rotating black holes and that the MAD relation in Eq.~(\ref{eq:magn_flux_MAD}) equally applies, we can assess the effects of spin on the energy spectrum and particle acceleration in magnetospheric current sheets. For example, the magnetic field intensity in the upstream region $B_0$, the synchrotron limited Lorentz factor $\gamma_{\rm rad}$, and consequently, the synchrotron luminosity of free pairs are all influenced by $f(a_{\rm s})$, which is defined as $1+\sqrt{1-a^2_{\rm s}}$ (see Eqs.~\ref{eq:magn_field}, \ref{eq:gamma_rad}, and \ref{eq:nLn_syn}). Given that $f$ ranges from 1 to 2 for maximally spinning and non-rotating black holes, respectively, the MeV peak synchrotron luminosity could be lower by a factor of  10 for a black hole with $a_{\rm s} \sim 0.1$. Additionally, the highest energy that protons can attain in the acceleration zone scales as $f^{-2}(a_{\rm s})$, with smaller spins corresponding to lower proton energies.  

For parameters relevant to M87* we showed that the trapped pair population cools due to synchrotron losses to $\gamma \ll \gamma_{\rm inj} \simeq \sigma_{\rm e}$ (see Eq.~\ref{eq:syn_cool}). Nonetheless, our model is based on the results of 3D PIC simulations without radiative losses. Synchrotron losses of plasma trapped in flux tubes can lead to reduced internal pressure, thus decreasing the transverse size of plasmoid structures in the reconnection region, as demonstrated in 3D radiative PIC simulations \cite{2023ApJ...959..122C}. In this case, the acceleration timescale of free particles, or the escape timescales of free particles from the region of active acceleration could differ from those adopted here. Moreover, the ratio $\xi$ between the free and the trapped population might be affected. To robustly address this issue, a detailed investigation of the free phase of acceleration in the regime of strong synchrotron cooling (i.e., $\gamma_{\rm cool}^{\rm syn} \ll \sigma_{\rm e} \ll \gamma_{\rm rad}$) is needed.

The numerical calculations of the non-thermal emission were performed in a single-zone framework (see Sec.~\ref{sub_sec:Numerical_appr}). According to this, all particle populations (free and trapped) occupy the same volume wherein the magnetic field and photon fields are homogeneous. In reality, the current sheet is a much more complex system, with trapped particles residing mainly in flux tubes and free particles moving above and below the current sheet, sampling the fields in the upstream region close to the sheet. Pairs entering the free phase are doing quasiperiodic deflections between the two sides of the reconnection layer with their distance from the midplane being $<0.1 l$ \cite{Zhang_2021}. Therefore in zeroth approximation, we can assume that the system is evolving in the same one-zone blob and the photon fields produced by the distinct pair populations coexist within the same region. 

We estimated the maximum energy that protons can achieve in magnetospheric current sheets, assuming the same rate of acceleration as for the free electrons. To make further predictions for the electromagnetic (e.g., proton synchrotron radiation) and neutrino signatures of the relativistic hadronic population, one would have to know the number density of protons in the magnetospheric region and the shape of their distribution. While it is unlikely that the plasma in the magnetospheric region of an accreting black hole consists of electrons and protons, it is possible for a small number of protons (compared to the pairs) to be present. Specifically, in our model, we have operated under the assumption that the upstream magnetization is governed by the pair number density; in other words, the rest mass energy density of the plasma is controlled by the pairs. Let us consider a proton population and adopt a power-law description for their number density distribution, represented as $n_{\rm p}(\gamma)=k_{\rm p} \gamma^{-p}$, where $k_{\rm p}$ and $p$ denote the normalization and power-law index respectively. Assuming that the power-law index would be $p\sim 1$, and the distribution would extend up to $\gamma^{\rm (syn)}_{\rm p,\max}$ (as indicated in Eq.~\ref{eq:gmax-syn}). We can then set an upper limit on the number density of protons in the region by considering that the energy density in relativistic protons cannot exceed the magnetic energy density available for dissipation. This leads to the following constraint on the proton density. We can demonstrate that by equating the magnetic field energy density with the proton energy density, the ratio of pairs to protons density in this system becomes $n_{\rm p}/n_{\rm e^{\pm}} < 0.25 \cdot 10^{-7} \sigma_{\rm e,4}(18.4+\ln(\gamma^{\rm (syn)}_{\rm p,\max,8}))^{-1}/\gamma^{\rm (syn)}_{\rm p,\max,8}$, where we normalized the maximum proton Lorentz factor to $10^8$ (motivated by figure~\ref{fig:E_p_max}). Consequently, if protons were accelerated into a hard power-law up to the radiation-limited Lorentz factor, their number density should be negligible compared to the pair density.

Throughout this work, we have assumed that the magnetization in the upstream region remains below the synchrotron radiation-limited Lorentz factor of pairs $\gamma_{\rm rad}$ ($\sigma_{\rm e} < \gamma_{\rm rad}$), in which case the scenario of particle acceleration during the free phase can operate. Recently \cite{2023ApJ...943L..29H} have discussed the production of TeV flares from pairs created via $\gamma \gamma$ pair production, in the current sheets of M87*. Despite the similarities to our work, there is a major difference regarding the regime of interest; Ref.~\cite{2023ApJ...943L..29H} focuses on cases where the initial magnetization of the upstream plasma (i.e. before pair enrichment) is much higher than $\gamma_{\rm rad}$ because the magnetospheric region is assumed to have a pair density equal to the Goldreich-Julian value (i.e. very low-density plasma region threaded by strong magnetic fields). In this regime, particles are mainly accelerated in the downstream reconnection region, with acceleration occurring predominantly in regions where the magnetic field strength is less than the electric field strength, as detailed in \cite{Sironi_2014}. There, particle Lorentz factors may exceed $\gamma_{\rm rad}$ due to anisotropic effects (i.e. small pitch angles) and potentially attain energies up to a few $\sigma_{\rm e} m_{\rm e}c^2$, contingent upon the strength of synchrotron cooling. In this scenario, $\gamma \gamma$ pair production takes place on synchrotron photons emitted by a single particle population. The calculations of multi-wavelength spectra are then performed assuming a steady state where $\sigma_{\rm e}$ is lower than its initial value as a result of pair enrichment. However, the pair density is calculated using optical depth arguments and does not incorporate dynamic effects. This might underestimate the total pair enrichment as shown in Appendix \ref{app:sigma}. 

In this work, we have operated under the assumption of a one-zone approach, wherein both pair populations $N^{\rm free}$ and $N^{\rm trap}$ occupy the same spatial volume and emit non-thermal photons isotropically as discussed in Section \ref{sub_sec:Numerical_appr}. However, to better describe real-world conditions, future investigations should aim to model pair production in scenarios where photons emitted by different pair populations have different geometrical distributions. Moreover, a more accurate calculation of pair production should incorporate interaction angles of photons and account for the production of pairs at different distances from the current sheet. Such an approach would offer a more nuanced understanding of secondary pair production rates and pair enrichment of the current sheet. Additionally, incorporating anisotropic effects into pair distributions is crucial, especially when pairs are accelerated above $\gamma_{\rm rad}$ (i.e. $\gamma_{\rm rad}<\sigma_{\rm e}$), as it can influence cooling processes \cite{Comisso_2023} and the resulting spectral characteristics. Finally in \ref{sub_sec:UHECR} we discussed the maximum energy that a single proton can achieve in the current sheet of M87*. It is essential to integrate realistic proton distributions and to delve into the intricate dynamics of proton acceleration in the context of 3D reconnection of pair-proton plasmas. An understanding of whether the free phase of acceleration applies to protons and the factors governing the broken power law distributions observed in recent simulations \cite{2023ApJ...959..122C} is essential. Finally, radiation losses as shown may play a role in the formation of the distribution. All these are essential to quantify the expected photon and neutrino emission arising from a proton population within the current sheet.

\section{Conclusions}
We presented a model of particle acceleration at current sheets formed in the magnetospheric region of a rotating SMBH. The particle distribution in the reconnection region consists of two distinct populations that are coined  ``free'' and ``trapped''. Free electrons and positrons are accelerated while meandering between both sides of a transient magnetospheric current sheet and can produce synchrotron-powered MeV flares. The luminosity of such flares, which is independent of the upstream plasma magnetization, is about 20 percent of the Blandford-Znajek power, and the duration of the flares is determined by the lifetime of the current sheets. Trapped electrons and positrons produce lower energy synchrotron radiation, which can extend down to a few MeV energies, and can produce X-ray counterparts to the MeV flares. Secondary pairs created via the attenuation of $\gtrsim$~MeV photons by lower energy synchrotron photons may enrich the plasma, thus reducing the initial upstream magnetization by a factor of 100 within the lifetime of the current sheet. Low-energy synchrotron photons from trapped pairs are important targets for pion production and can limit proton acceleration to EeV energies for parameters relevant to M87*.

\bibliographystyle{JHEP}
\bibliography{paper}

\providecommand{\href}[2]{#2}\begingroup\raggedright\begin{thebibliography}{10}

\bibitem{Akiyama_2019}
T.E.H.T.~Collaboration, K.~Akiyama, A.~Alberdi, W.~Alef, K.~Asada, R.~Azulay et~al., \emph{First m87 event horizon telescope results. vi. the shadow and mass of the central black hole}, \href{https://doi.org/10.3847/2041-8213/ab1141}{\emph{The Astrophysical Journal Letters} {\bfseries 875} (2019) L6}.

\bibitem{2009Sci...325..444A}
V.A.~{Acciari}, E.~{Aliu}, T.~{Arlen}, M.~{Bautista}, M.~{Beilicke}, W.~{Benbow} et~al., \emph{{Radio Imaging of the Very-High-Energy {\ensuremath{\gamma}}-Ray Emission Region in the Central Engine of a Radio Galaxy}}, \href{https://doi.org/10.1126/science.1175406}{\emph{Science} {\bfseries 325} (2009) 444} [\href{https://arxiv.org/abs/0908.0511}{{\ttfamily 0908.0511}}].

\bibitem{Harris_2009}
D.E.~Harris, C.C.~Cheung, Łukasz Stawarz, J.A.~Biretta and E.S.~Perlman, \emph{Variability timescales in the m87 jet: Signatures of e2 losses, discovery of a quasi period in hst-1, and the site of tev flaring}, \href{https://doi.org/10.1088/0004-637X/699/1/305}{\emph{The Astrophysical Journal} {\bfseries 699} (2009) 305}.

\bibitem{Aliu_2012}
E.~{Aliu}, T.~{Arlen}, T.~{Aune}, M.~{Beilicke}, W.~{Benbow}, A.~{Bouvier} et~al., \emph{{VERITAS Observations of Day-scale Flaring of M 87 in 2010 April}}, \href{https://doi.org/10.1088/0004-637X/746/2/141}{\emph{\apj} {\bfseries 746} (2012) 141} [\href{https://arxiv.org/abs/1112.4518}{{\ttfamily 1112.4518}}].

\bibitem{2016Natur.531..476H}
{HESS Collaboration}, A.~{Abramowski}, F.~{Aharonian}, F.A.~{Benkhali}, A.G.~{Akhperjanian}, E.O.~{Ang{\"u}ner} et~al., \emph{{Acceleration of petaelectronvolt protons in the Galactic Centre}}, \href{https://doi.org/10.1038/nature17147}{\emph{Nature} {\bfseries 531} (2016) 476} [\href{https://arxiv.org/abs/1603.07730}{{\ttfamily 1603.07730}}].

\bibitem{Rieger_2022}
F.M.~Rieger, \emph{Active galactic nuclei as potential sources of ultra-high energy cosmic rays}, \href{https://doi.org/10.3390/universe8110607}{\emph{Universe} {\bfseries 8} (2022) }.

\bibitem{Akiyama_2021}
T.E.H.T.~Collaboration, K.~Akiyama, J.C.~Algaba, A.~Alberdi, W.~Alef, R.~Anantua et~al., \emph{First m87 event horizon telescope results. viii. magnetic field structure near the event horizon}, \href{https://doi.org/10.3847/2041-8213/abe4de}{\emph{The Astrophysical Journal Letters} {\bfseries 910} (2021) L13}.

\bibitem{2022ApJ...924L..32R}
B.~{Ripperda}, M.~{Liska}, K.~{Chatterjee}, G.~{Musoke}, A.A.~{Philippov}, S.B.~{Markoff} et~al., \emph{{Black Hole Flares: Ejection of Accreted Magnetic Flux through 3D Plasmoid-mediated Reconnection}}, \href{https://doi.org/10.3847/2041-8213/ac46a1}{\emph{ApJ Lett.} {\bfseries 924} (2022) L32} [\href{https://arxiv.org/abs/2109.15115}{{\ttfamily 2109.15115}}].

\bibitem{Lyutikov_2003}
M.~Lyutikov and D.~Uzdensky, \emph{Dynamics of relativistic reconnection}, \href{https://doi.org/10.1086/374808}{\emph{The Astrophysical Journal} {\bfseries 589} (2003) 893}.

\bibitem{Lyubarsky_2005}
Y.E.~Lyubarsky, \emph{{On the relativistic magnetic reconnection}}, \href{https://doi.org/10.1111/j.1365-2966.2005.08767.x}{\emph{Monthly Notices of the Royal Astronomical Society} {\bfseries 358} (2005) 113} [\href{https://arxiv.org/abs/https://academic.oup.com/mnras/article-pdf/358/1/113/3473070/358-1-113.pdf}{{\ttfamily https://academic.oup.com/mnras/article-pdf/358/1/113/3473070/358-1-113.pdf}}].

\bibitem{2016ApJ...816L...8W}
G.R.~{Werner}, D.A.~{Uzdensky}, B.~{Cerutti}, K.~{Nalewajko} and M.C.~{Begelman}, \emph{{The Extent of Power-law Energy Spectra in Collisionless Relativistic Magnetic Reconnection in Pair Plasmas}}, \href{https://doi.org/10.3847/2041-8205/816/1/L8}{\emph{ApJL} {\bfseries 816} (2016) L8} [\href{https://arxiv.org/abs/1409.8262}{{\ttfamily 1409.8262}}].

\bibitem{Sironi_2014}
L.~Sironi and A.~Spitkovsky, \emph{Relativistic reconnection: An efficient source of non-thermal particles}, \href{https://doi.org/10.1088/2041-8205/783/1/L21}{\emph{The Astrophysical Journal Letters} {\bfseries 783} (2014) L21}.

\bibitem{2022PhRvL.128n5102S}
L.~{Sironi}, \emph{{Nonideal Fields Solve the Injection Problem in Relativistic Reconnection}}, \href{https://doi.org/10.1103/PhysRevLett.128.145102}{\emph{PRL} {\bfseries 128} (2022) 145102} [\href{https://arxiv.org/abs/2203.04342}{{\ttfamily 2203.04342}}].

\bibitem{2018MNRAS.481.5687P}
M.~{Petropoulou} and L.~{Sironi}, \emph{{The steady growth of the high-energy spectral cut-off in relativistic magnetic reconnection}}, \href{https://doi.org/10.1093/mnras/sty2702}{\emph{\mnras} {\bfseries 481} (2018) 5687} [\href{https://arxiv.org/abs/1808.00966}{{\ttfamily 1808.00966}}].

\bibitem{2021ApJ...912...48H}
H.~{Hakobyan}, M.~{Petropoulou}, A.~{Spitkovsky} and L.~{Sironi}, \emph{{Secondary Energization in Compressing Plasmoids during Magnetic Reconnection}}, \href{https://doi.org/10.3847/1538-4357/abedac}{\emph{\apj} {\bfseries 912} (2021) 48} [\href{https://arxiv.org/abs/2006.12530}{{\ttfamily 2006.12530}}].

\bibitem{Zhang_2021}
H.~Zhang, L.~Sironi and D.~Giannios, \emph{Fast particle acceleration in three-dimensional relativistic reconnection}, \href{https://doi.org/10.3847/1538-4357/ac2e08}{\emph{The Astrophysical Journal} {\bfseries 922} (2021) 261}.

\bibitem{2023ApJ...956L..36Z}
H.~{Zhang}, L.~{Sironi}, D.~{Giannios} and M.~{Petropoulou}, \emph{{The Origin of Power-law Spectra in Relativistic Magnetic Reconnection}}, \href{https://doi.org/10.3847/2041-8213/acfe7c}{\emph{ApJL} {\bfseries 956} (2023) L36} [\href{https://arxiv.org/abs/2302.12269}{{\ttfamily 2302.12269}}].

\bibitem{2003PASJ...55L..69N}
R.~{Narayan}, I.V.~{Igumenshchev} and M.A.~{Abramowicz}, \emph{{Magnetically Arrested Disk: an Energetically Efficient Accretion Flow}}, \href{https://doi.org/10.1093/pasj/55.6.L69}{\emph{PASJ} {\bfseries 55} (2003) L69} [\href{https://arxiv.org/abs/astro-ph/0305029}{{\ttfamily astro-ph/0305029}}].

\bibitem{2011MNRAS.418L..79T}
A.~{Tchekhovskoy}, R.~{Narayan} and J.C.~{McKinney}, \emph{{Efficient generation of jets from magnetically arrested accretion on a rapidly spinning black hole}}, \href{https://doi.org/10.1111/j.1745-3933.2011.01147.x}{\emph{\mnras} {\bfseries 418} (2011) L79} [\href{https://arxiv.org/abs/1108.0412}{{\ttfamily 1108.0412}}].

\bibitem{2002apa..book.....F}
J.~{Frank}, A.~{King} and D.J.~{Raine}, \emph{{Accretion Power in Astrophysics: Third Edition}} (2002).

\bibitem{2021ApJ...922..261Z}
H.~{Zhang}, L.~{Sironi} and D.~{Giannios}, \emph{{Fast Particle Acceleration in Three-dimensional Relativistic Reconnection}}, \href{https://doi.org/10.3847/1538-4357/ac2e08}{\emph{ApJ} {\bfseries 922} (2021) 261} [\href{https://arxiv.org/abs/2105.00009}{{\ttfamily 2105.00009}}].

\bibitem{2023PhRvL.130r9501G}
F.~{Guo}, X.~{Li}, O.~{French}, Q.~{Zhang}, W.~{Daughton}, Y.-H.~{Liu} et~al., \emph{{Comment on ``Nonideal Fields Solve the Injection Problem in Relativistic Reconnection''}}, \href{https://doi.org/10.1103/PhysRevLett.130.189501}{\emph{PRL} {\bfseries 130} (2023) 189501} [\href{https://arxiv.org/abs/2208.03435}{{\ttfamily 2208.03435}}].

\bibitem{1998A&A...333..452K}
J.G.~{Kirk}, F.M.~{Rieger} and A.~{Mastichiadis}, \emph{{Particle acceleration and synchrotron emission in blazar jets}}, \href{https://doi.org/10.48550/arXiv.astro-ph/9801265}{\emph{A\&A} {\bfseries 333} (1998) 452} [\href{https://arxiv.org/abs/astro-ph/9801265}{{\ttfamily astro-ph/9801265}}].

\bibitem{1977MNRAS.179..433B}
R.D.~{Blandford} and R.L.~{Znajek}, \emph{{Electromagnetic extraction of energy from Kerr black holes.}}, \href{https://doi.org/10.1093/mnras/179.3.433}{\emph{\mnras} {\bfseries 179} (1977) 433}.

\bibitem{1990MNRAS.245..453C}
P.S.~{Coppi} and R.D.~{Blandford}, \emph{{Reaction rates and energy distributions for elementary processes in relativistic pair plasmas}}, \href{https://doi.org/10.1093/mnras/245.3.453}{\emph{\mnras} {\bfseries 245} (1990) 453}.

\bibitem{10.1093/mnras/stw1620}
L.~Sironi, D.~Giannios and M.~Petropoulou, \emph{{Plasmoids in relativistic reconnection, from birth to adulthood: first they grow, then they go}}, \href{https://doi.org/10.1093/mnras/stw1620}{\emph{Monthly Notices of the Royal Astronomical Society} {\bfseries 462} (2016) 48} [\href{https://arxiv.org/abs/https://academic.oup.com/mnras/article-pdf/462/1/48/18755738/stw1620.pdf}{{\ttfamily https://academic.oup.com/mnras/article-pdf/462/1/48/18755738/stw1620.pdf}}].

\bibitem{2001ApJ...562L..63Z}
S.~{Zenitani} and M.~{Hoshino}, \emph{{The Generation of Nonthermal Particles in the Relativistic Magnetic Reconnection of Pair Plasmas}}, \href{https://doi.org/10.1086/337972}{\emph{ApJL} {\bfseries 562} (2001) L63} [\href{https://arxiv.org/abs/1402.7139}{{\ttfamily 1402.7139}}].

\bibitem{2008ApJ...682.1436L}
Y.~{Lyubarsky} and M.~{Liverts}, \emph{{Particle Acceleration in the Driven Relativistic Reconnection}}, \href{https://doi.org/10.1086/589640}{\emph{\apj} {\bfseries 682} (2008) 1436} [\href{https://arxiv.org/abs/0805.0085}{{\ttfamily 0805.0085}}].

\bibitem{Inoue_Takahara}
S.~{Inoue} and F.~{Takahara}, \emph{{Electron Acceleration and Gamma-Ray Emission from Blazars}}, \href{https://doi.org/10.1086/177270}{\emph{ApJ} {\bfseries 463} (1996) 555}.

\bibitem{2024A&A...683A.225S}
S.I.~{Stathopoulos}, M.~{Petropoulou}, G.~{Vasilopoulos} and A.~{Mastichiadis}, \emph{{LeHaMoC: A versatile time-dependent lepto-hadronic modeling code for high-energy astrophysical sources}}, \href{https://doi.org/10.1051/0004-6361/202347277}{\emph{A\&A} {\bfseries 683} (2024) A225} [\href{https://arxiv.org/abs/2308.06174}{{\ttfamily 2308.06174}}].

\bibitem{2023ApJ...959..122C}
A.~{Chernoglazov}, H.~{Hakobyan} and A.~{Philippov}, \emph{{High-energy Radiation and Ion Acceleration in Three-dimensional Relativistic Magnetic Reconnection with Strong Synchrotron Cooling}}, \href{https://doi.org/10.3847/1538-4357/acffc6}{\emph{\apj} {\bfseries 959} (2023) 122} [\href{https://arxiv.org/abs/2305.02348}{{\ttfamily 2305.02348}}].

\bibitem{distance_m87}
{Bird, S.}, {Harris, W. E.}, {Blakeslee, J. P.} and {Flynn, C.}, \emph{The inner halo of m a first direct view of the red-giant population}, \href{https://doi.org/10.1051/0004-6361/201014876}{\emph{A\&A} {\bfseries 524} (2010) A71}.

\bibitem{2011ApJ...735....9M}
M.~{Mo{\'s}cibrodzka}, C.F.~{Gammie}, J.C.~{Dolence} and H.~{Shiokawa}, \emph{{Pair Production in Low-luminosity Galactic Nuclei}}, \href{https://doi.org/10.1088/0004-637X/735/1/9}{\emph{\apj} {\bfseries 735} (2011) 9} [\href{https://arxiv.org/abs/1104.2042}{{\ttfamily 1104.2042}}].

\bibitem{refId0}
{Baes, M.}, {Clemens, M.}, {Xilouris, E. M.}, {Fritz, J.}, {Cotton, W. D.}, {Davies, J. I.} et~al., \emph{The herschel virgo cluster survey * - vi. the far-infrared view of m}, \href{https://doi.org/10.1051/0004-6361/201014555}{\emph{A\&A} {\bfseries 518} (2010) L53}.

\bibitem{Algaba_2021}
T.E.M.S.W.~Group, J.C.~Algaba, J.~Anczarski, K.~Asada, M.~Baloković, S.~Chandra et~al., \emph{Broadband multi-wavelength properties of m87 during the 2017 event horizon telescope campaign}, \href{https://doi.org/10.3847/2041-8213/abef71}{\emph{The Astrophysical Journal Letters} {\bfseries 911} (2021) L11}.

\bibitem{2023RAA....23f5018C}
Y.-L.~{Cheng}, F.~{Xiang}, H.~{Yu}, S.-M.~{Jia}, X.-H.~{Li}, C.-K.~{Li} et~al., \emph{{The Year-scale X-Ray Variations in the Core of M87}}, \href{https://doi.org/10.1088/1674-4527/accb7a}{\emph{Research in Astronomy and Astrophysics} {\bfseries 23} (2023) 065018} [\href{https://arxiv.org/abs/2303.12353}{{\ttfamily 2303.12353}}].

\bibitem{2022ApJ...937L..34K}
S.S.~{Kimura}, K.~{Toma}, H.~{Noda} and K.~{Hada}, \emph{{Magnetic Reconnection in Black Hole Magnetospheres: Lepton Loading into Jets, Superluminal Radio Blobs, and Multiwavelength Flares}}, \href{https://doi.org/10.3847/2041-8213/ac8d5a}{\emph{ApJL} {\bfseries 937} (2022) L34} [\href{https://arxiv.org/abs/2208.01882}{{\ttfamily 2208.01882}}].

\bibitem{2023ApJ...944..173C}
A.Y.~{Chen}, D.~{Uzdensky} and J.~{Dexter}, \emph{{Synchrotron Pair Production Equilibrium in Relativistic Magnetic Reconnection}}, \href{https://doi.org/10.3847/1538-4357/acb68a}{\emph{\apj} {\bfseries 944} (2023) 173} [\href{https://arxiv.org/abs/2209.03249}{{\ttfamily 2209.03249}}].

\bibitem{2022JATIS...8d4003C}
R.~{Caputo}, M.~{Ajello}, C.A.~{Kierans}, J.S.~{Perkins}, J.L.~{Racusin}, L.~{Baldini} et~al., \emph{{All-sky Medium Energy Gamma-ray Observatory eXplorer mission concept}}, \href{https://doi.org/10.1117/1.JATIS.8.4.044003}{\emph{Journal of Astronomical Telescopes, Instruments, and Systems} {\bfseries 8} (2022) 044003} [\href{https://arxiv.org/abs/2208.04990}{{\ttfamily 2208.04990}}].

\bibitem{SraA_EHT}
E.H.T.~Collaboration, K.~Akiyama, A.~Alberdi, W.~Alef, J.C.~Algaba, R.~Anantua et~al., \emph{First sagittarius a* event horizon telescope results. i. the shadow of the supermassive black hole in the center of the milky way}, \href{https://doi.org/10.3847/2041-8213/ac6674}{\emph{The Astrophysical Journal Letters} {\bfseries 930} (2022) L12}.

\bibitem{2023ApJ...943L..29H}
H.~{Hakobyan}, B.~{Ripperda} and A.A.~{Philippov}, \emph{{Radiative Reconnection-powered TeV Flares from the Black Hole Magnetosphere in M87}}, \href{https://doi.org/10.3847/2041-8213/acb264}{\emph{ApJL} {\bfseries 943} (2023) L29} [\href{https://arxiv.org/abs/2209.02105}{{\ttfamily 2209.02105}}].

\bibitem{Comisso_2023}
L.~Comisso and B.~Jiang, \emph{Pitch-angle anisotropy imprinted by relativistic magnetic reconnection}, \href{https://doi.org/10.3847/1538-4357/ad1241}{\emph{The Astrophysical Journal} {\bfseries 959} (2023) 137}.

\bibitem{1990ApJ...362...38B}
M.C.~{Begelman}, B.~{Rudak} and M.~{Sikora}, \emph{{Consequences of Relativistic Proton Injection in Active Galactic Nuclei}}, \href{https://doi.org/10.1086/169241}{\emph{\apj} {\bfseries 362} (1990) 38}.

\bibitem{1992ApJ...400..181C}
M.J.~{Chodorowski}, A.A.~{Zdziarski} and M.~{Sikora}, \emph{{Reaction Rate and Energy-Loss Rate for Photopair Production by Relativistic Nuclei}}, \href{https://doi.org/10.1086/171984}{\emph{\apj} {\bfseries 400} (1992) 181}.

\end{thebibliography}\endgroup

\acknowledgments
We thank H. Hakobyan for his insightful comments and discussions. S.I.S. and M.P. acknowledge support from the Hellenic Foundation for Research and Innovation (H.F.R.I.) under the ``2nd call for H.F.R.I. Research Projects to support Faculty members and Researchers'' through the project UNTRAPHOB (Project ID 3013). L.S. acknowledges support from DoE Early Career Award DE-SC0023015. This work was supported by a grant from the Simons Foundation (MP-SCMPS-00001470) to L.S., and facilitated by Multimessenger Plasma Physics Center (MPPC).

\appendix 
\section{Determination of $\zeta$}\label{app:zeta}

We can infer $\zeta$ using results from recent 3D PIC simulations of relativistic reconnection in pair plasmas \cite{2023ApJ...956L..36Z}. It was shown (see Fig. 6 in Ref.~\cite{2023ApJ...956L..36Z}) that the number of pairs with $\gamma \sim \gamma_{\rm inj}$ that have experienced at least one free phase of acceleration is $\xi \sim 0.3$ of the total number of pairs in the system $N_{\rm tot}\equiv N^{\rm free}+N^{\rm trap}_{\rm fr}+N^{\rm trap}_{\rm X}$ (in steady state). These simulations were non-radiative (i.e. cooling of pairs was not included).

The ratio $\xi$ can then be expressed as

\begin{equation}
\xi \equiv \frac{\Bigg[\gamma N^{\rm free}(\gamma)+\gamma N^{\rm trap}_{\rm fr}(\gamma)\Bigg]_{\gamma=\gamma_{\rm inj}}}{\Bigg[\gamma N_{\rm tot}(\gamma)\Bigg]_{\gamma=\gamma_{\rm inj}}},
\label{eq:xi}
\end{equation}
where the steady-state pair distributions (in the absence of radiative cooling) read,

\begin{equation}
N^{\rm free} (\gamma) = \zeta Q^{\rm tot}_{\rm e,inj}\gamma^{-1}t_{\rm acc}(\gamma), \, \gamma \ge \gamma_{\rm inj}
\label{eq:pair_distribution_free}
\end{equation}

\begin{equation}
N^{\rm trap}_{\rm fr}(\gamma) = \zeta Q^{\rm tot}_{\rm e,inj}\gamma^{-2}\gamma_{\rm inj} t_{\rm adv}, \,  \gamma \ge \gamma_{\rm inj}
\label{eq:pair_distribution_free_trapped}
\end{equation}
and
\begin{equation}
N^{\rm trap}_{\rm X}(\gamma) = (1-\zeta)\frac{Q^{\rm tot}_{\rm e,inj}}{\ln(\sigma_{\rm e})}\gamma^{-1}t_{\rm adv}, \,  \gamma \le \gamma_{\rm inj}.
\label{eq:pair_distribution_X_trapped}
\end{equation}

By substituting Eqs.~(\ref{eq:pair_distribution_free}) - (\ref{eq:pair_distribution_X_trapped}) into Eq.~(\ref{eq:xi}) we find that 

\begin{equation}
\xi = \frac{\zeta\big(1+\frac{t_{\rm acc}(\gamma_{\rm inj})}{t_{\rm adv}}\big)}{\zeta \big(1+\frac{t_{\rm acc}(\gamma_{\rm inj})}{t_{\rm adv}}\big) + (1-\zeta)\frac{1}{\ln(\sigma_{\rm e})}}.
\label{eq:xi_zeta}
\end{equation}

The ratio of the two timescales at $\gamma_{\rm inj} \sim \sigma_{\rm e}$ is given by 
\begin{equation}
\frac{t_{\rm acc}}{t_{\rm adv}} \simeq 10^{-12.7} \sigma_{\rm e} \, \frac{(\eta_{\rm c,-1}\dot{m}_{-5}M_9^{-1})^{1/2}f^2(a_{\rm s})}{\eta_{\rm rec,-1}} \ll 1,
\label{eq:timescale_ratio}
\end{equation}

and Eq.~(\ref{eq:xi_zeta}) simplifies to
\begin{equation}
\zeta \approx \big[1 + \ln(\sigma_{\rm e}) \big(\frac{1}{\xi} -1 \big) \big]^{-1},
\end{equation}
which depends only logarithmically on $\sigma_{\rm e}$. For $\sigma_{\rm e} \in (10, 10^6)$, which is our range of interest, we find an average value of $0.06$ that we adopt throughout this work.

\section{Optical depth to $\gamma \gamma$ pair production}\label{app:tau_gg}

In Sec.~\ref{sec:free} we estimated the optical depth for $\gamma \gamma$ pair creation considering as target photons only those produced by the synchrotron radiation of free pairs and showed that $\tau_{\gamma \gamma} \sim \mathcal{O}(10^{-4})$ (see Eq.~\ref{eq:tau_gg_free} for the dependence on model parameters). Thermal disk photons are not relevant targets for the attenuation of 10~MeV photons. In the following, we estimate the optical depth after accounting for the synchrotron photons from the trapped pairs.

The photon spectrum of the target photons for $\gamma \gamma$ pair creation depends on $\sigma_{\rm e}$. Let $\epsilon_{\rm syn}\sim 10$~MeV denote the peak synchrotron frequency of the free pair population (see Eq.~\ref{eq:nu_c}), and $\epsilon_{\rm l}$ denote the energy of target photons interacting with the $\epsilon_{\rm syn}$ $\gamma$-rays close to the pair production threshold (since the cross-section of the interaction peaks near the threshold). The majority of these target photons are produced by the $N^{\rm trap}_{\rm fr}$ pairs which are injected with $\gamma_{\rm inj}\sim \sigma_{\rm e}<\gamma<\gamma_{\rm rad}$, and subsequently cool to lower energies. In general $N^{\rm trap}_{\rm fr}$ can be approximated by a broken power law with slopes $\sim$ -2 and $\sim$-3 below and above $\sigma_{\rm e}$ respectively (see Eq.~\ref{eq:trapped_pairs_dis_approx}). The corresponding synchrotron spectrum will then be $L_{\rm syn}(\epsilon)\propto \epsilon^{-1/2}$ and $\propto \epsilon^{-1}$ below and above $\epsilon_{\rm syn}(\sigma_{\rm e})$ respectively. In the context of M87* the luminosity of target photons can be approximated as $L_{\rm l}\approx L_{\rm syn}(\epsilon_{\rm l}/\epsilon_{\rm syn}(\sigma_{\rm e}))^{1/2}$ if $\epsilon_{\rm syn}(\sigma_{\rm e})>\epsilon_{\rm l} \Leftrightarrow \sigma_{\rm e}>10^{5.2}$, or  $L_{\rm l}\approx L_{\rm syn}$ otherwise.
 
The optical depth for $\gamma \gamma$ pair creation in these two scenarios can then be written as 

\begin{equation}
\tau_{\rm \gamma \gamma} \simeq 
\begin{cases} 
      0.007\dot{m}_{-5}\mathcal{R}_0 \frac{\zeta_{-1.2}\eta_{\rm rec,-1}^2}{f^4(a_{\rm s})}, & \sigma_{\rm e}<10^{5.2} \\
        0.004 \dot{m}^{3/4}_{-5}\mathcal{R}_0 \sigma_{\rm e,5.2}^{-1} \frac{\zeta_{-1.2}(M_9\eta_{\rm c,-1})^{1/4}\eta_{\rm rec,-1}^{3/2}}{f^3(a_{\rm s})}& \sigma_{\rm e}>10^{5.2} \\
   \end{cases}
\label{eq:tau_gg_trapped}
\end{equation}
assuming that $L_{\rm syn}$ is given by \ref{eq:nLn_syn}.

\begin{figure*}%
    \centering
\includegraphics[width=0.6\textwidth]{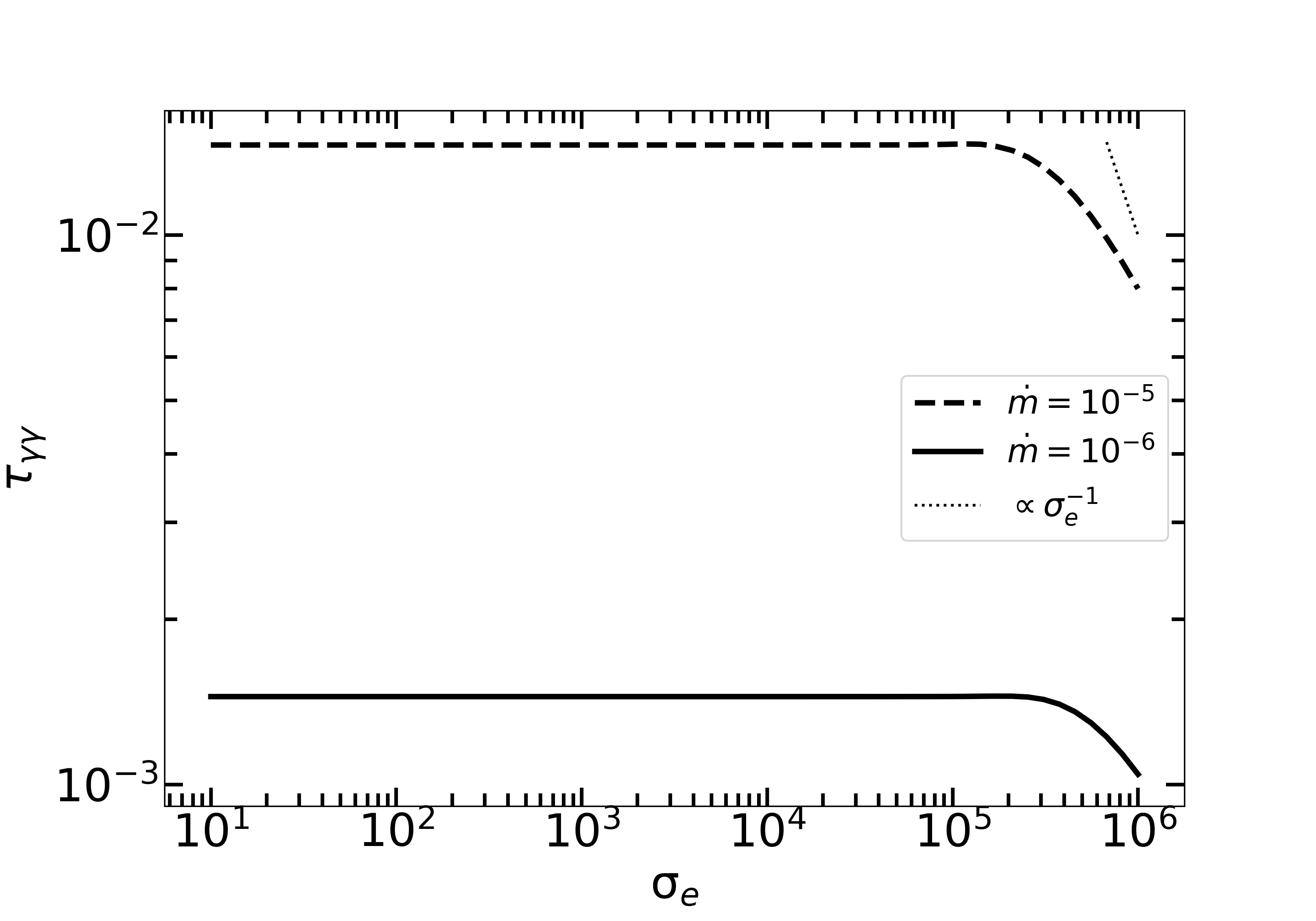} 
    \caption{Optical depth of pair creation for two values of $\dot{m}$ as a function of pair magnetization $\sigma_{\rm e}$. All other parameters are the same as in Table \ref{tab:param}}%
    \label{fig:tau_gg}%
\end{figure*}

In Fig. \ref{fig:tau_gg} we plot the optical depth computed numerically using {\tt LeHaMoC} for two values of $\dot{m}$ (all other parameters are the same as in Table~\ref{tab:param}), which agrees well with the analytical estimate. The inclusion of photons produced by trapped pairs as targets for $\gamma \gamma$ absorption increases the optical depth of the interaction and thus the pair creation rate within the system's volume. 

\section{Proton timescales}\label{app:times}
The proton synchrotron energy loss timescale is defined as,
\begin{equation}
t_{\rm syn, p} \approx\gamma_{\rm p}\Bigg(\frac{d\gamma_{\rm p}}{dt}\Bigg)_{\rm syn}^{-1}\simeq 3 \cdot 10^{-3}~{\rm s} \,  (\gamma_{\rm p,6}\dot{m}_{-5})^{-1}M_9\eta_{\rm c,-1}f^4(a_{\rm s})\Bigg(\frac{m_{\rm p}}{m_{\rm e}}\Bigg)^3 
\label{eq:syn_ts}
\end{equation}

The synchrotron radiation-limited proton Lorentz factor reads 
\begin{equation} 
\gamma_{\rm p, max}^{(\rm syn)} \simeq \sqrt{3}\cdot 10^6 \
M_9^{1/4} \eta_{\rm c,-1}^{-1/4} \dot{m}_{-5}^{-1/4} \eta_{\rm rec, -1}^{1/2} f(a_{\rm s}) \left(  \frac{m_{\rm p}}{m_{\rm e}} \right)
\label{eq:gmax-syn}
\end{equation}

The energy loss time scale of a high energy proton with Lorentz factor $\gamma_{\rm p}$ due to photopion production is given by \citep{1990ApJ...362...38B},

\begin{equation}
t_{\rm p\gamma,\pi}\approx\gamma_{\rm p}\Bigg(\frac{d\gamma_{\rm p}}{dt}\Bigg)_{\rm p\gamma,\pi}^{-1}\simeq \Bigg(\frac{c}{2\gamma_{\rm p}^2}\int_{\bar{\epsilon}_{\rm th}}^{\infty}d\bar{\epsilon}\sigma_{\rm p \pi}(\bar{\epsilon})k_{\rm p\pi}(\bar{\epsilon})\bar{\epsilon}\int_{\bar{\epsilon}/2\gamma_{\rm p}}^{\infty}d\epsilon\frac{n_{\gamma}(\epsilon)}{\epsilon^2}\Bigg)^{-1},
\label{eq:pg,pi_losses}
\end{equation}
where $\bar{\epsilon}_{\rm th}=145$  MeV, $n_{\gamma}(\epsilon)\equiv dN_{\gamma}/dVd\epsilon$ denotes the differential density of photons in the source frame of reference, while all barred quantities are measured in the proton's rest frame. The energy-dependent cross-section and the inelasticity (fraction of energy transferred per collision to secondaries) of the process are represented by $\sigma_{\rm p\pi}$ and $k_{\rm p\pi} $, respectively. 

The energy loss timescale for Bethe-Heitler pair production of a high energy proton with Lorentz factor $\gamma_{\rm p}$ is given by \citep{1992ApJ...400..181C} 

\begin{equation}
t_{\rm p\gamma, ee}\approx\gamma_{\rm p}\Bigg(\frac{d\gamma_{\rm p}}{dt}\Bigg)_{\rm p\gamma, ee}^{-1}\simeq \Bigg(\frac{3\sigma_{\rm T}\alpha_{\rm f} m_{\rm e}}{8\pi\gamma_{\rm p}}\int_{2}^{\infty}dk\ n_{\gamma}\bigg(\frac{k}{2\gamma_{\rm p}}\bigg)\frac{\phi(k)}{k^2}\Bigg)^{-1}
\label{eq:pg,BH_losses}
\end{equation}
where $\alpha_{\rm f}$ is the fine structure constant, $\sigma_{\rm T}$ is the Thomson cross-section, and $k=2\gamma_{\rm p}\epsilon/(m_{\rm e}c^2)$ is the photon energy in the proton's rest frame in units of the electron rest mass energy. 

\section{Time-dependent magnetization and pair enrichment} \label{app:sigma}
We observe that for certain combinations of $\dot{m}$ and $\sigma_{\rm e}$ the density of secondary pairs generated in the system through $\gamma \gamma$ absorption may exceed the primary density in a steady state (see Fig. \ref{fig:ratio_density_vs_sigma_e}). In order to assess whether pair creation influences the system dynamics, we conducted numerical calculations using the radiative code {\tt LeHaMoC}. In these numerical experiments, we allowed the magnetization $\sigma_{\rm e}$ of the system to vary over time. The recalculation of $\sigma_{\rm e}$ was done every $10 R_{\rm eff}/c$ in order to let the system approach its new steady state after each modification of the magnetization. 

First, we can calculate analytically the new magnetization after each pair enrichment using  Eq.~(\ref{eq:ratio_eq}). If we initiate the numerical calculation with magnetization $\sigma_0$, which corresponds to a number density $n_{\rm e^{\pm},0}$, then after $10 R_{\rm eff}/c$ when the system will be close to a steady state,  the total number density of pairs will be,
\begin{equation}
n_{\rm e^{\pm},1}=n_{\rm e^{\pm},0}(1+\eta_{\rm r}(\sigma_0))
\label{eq:n_e_update}
\end{equation}
where $\eta_{\rm r}\equiv n^{\rm sec}_{\rm \gamma \gamma}/n_{\rm e^{\pm}}$ which is a function of $\sigma_{\rm e}$ (see Eq. \ref{eq:ratio_eq}). Modifying the magnetization accordingly we find that,

\begin{equation}
\sigma_1=\sigma_0 \frac{n_{\rm e^{\pm},0}}{n_{\rm e^{\pm},1}}
\label{eq:sigma_e_update}
\end{equation}
By repeating the same procedure we find that the magnetization in the  step $i+1$ of the modification will be given by,

\begin{equation}
\sigma_{\rm i+1}= \frac{\sigma_0}{\prod_{\rm i=0}^{\rm i_{\max}} (1+\eta_{\rm r}(\sigma_{\rm i}))}.
\label{eq:sigma_e_i}
\end{equation}

\begin{figure*}%
    \centering
\includegraphics[width=0.49 \textwidth]{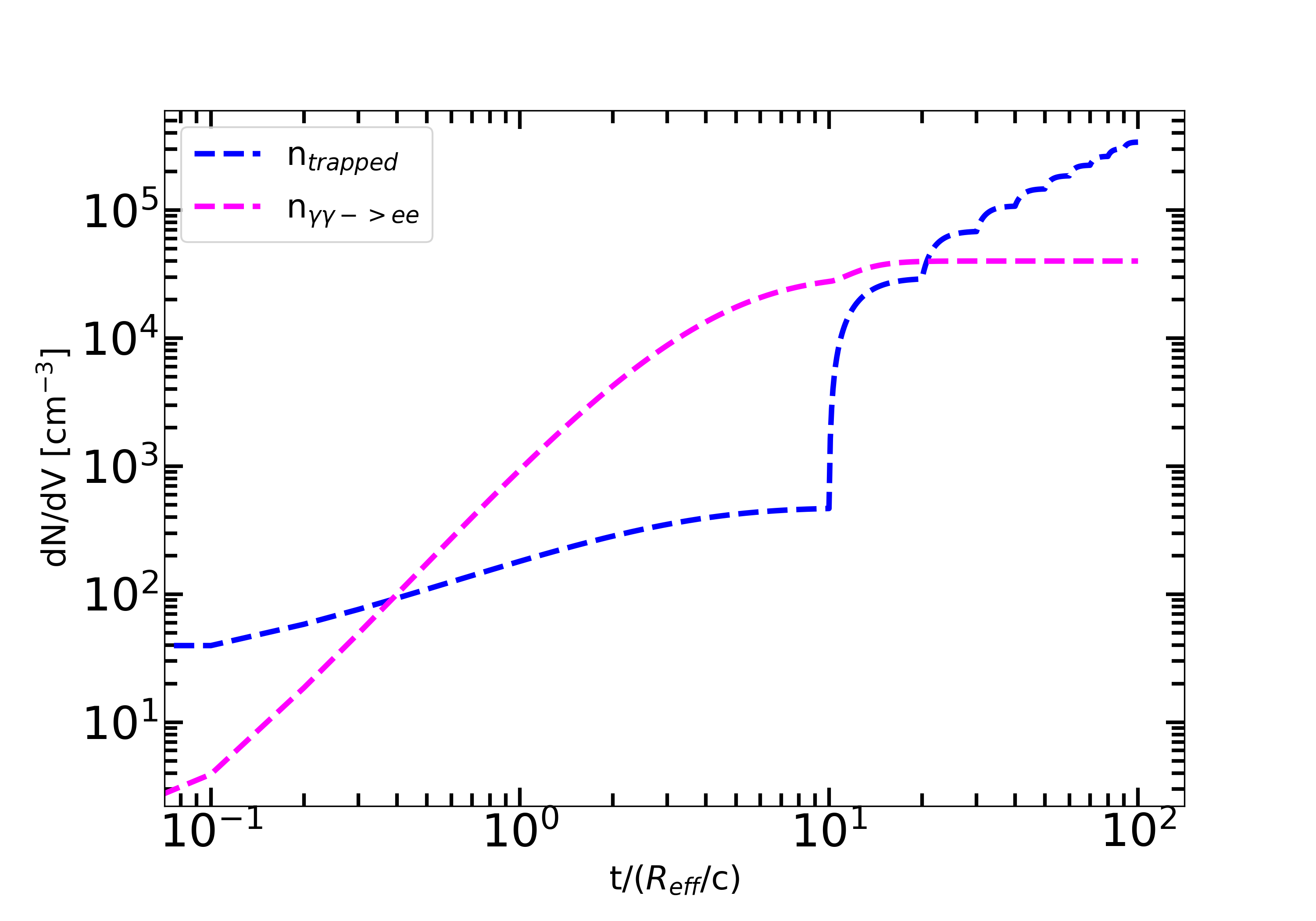}
\includegraphics[width=0.49 \textwidth]{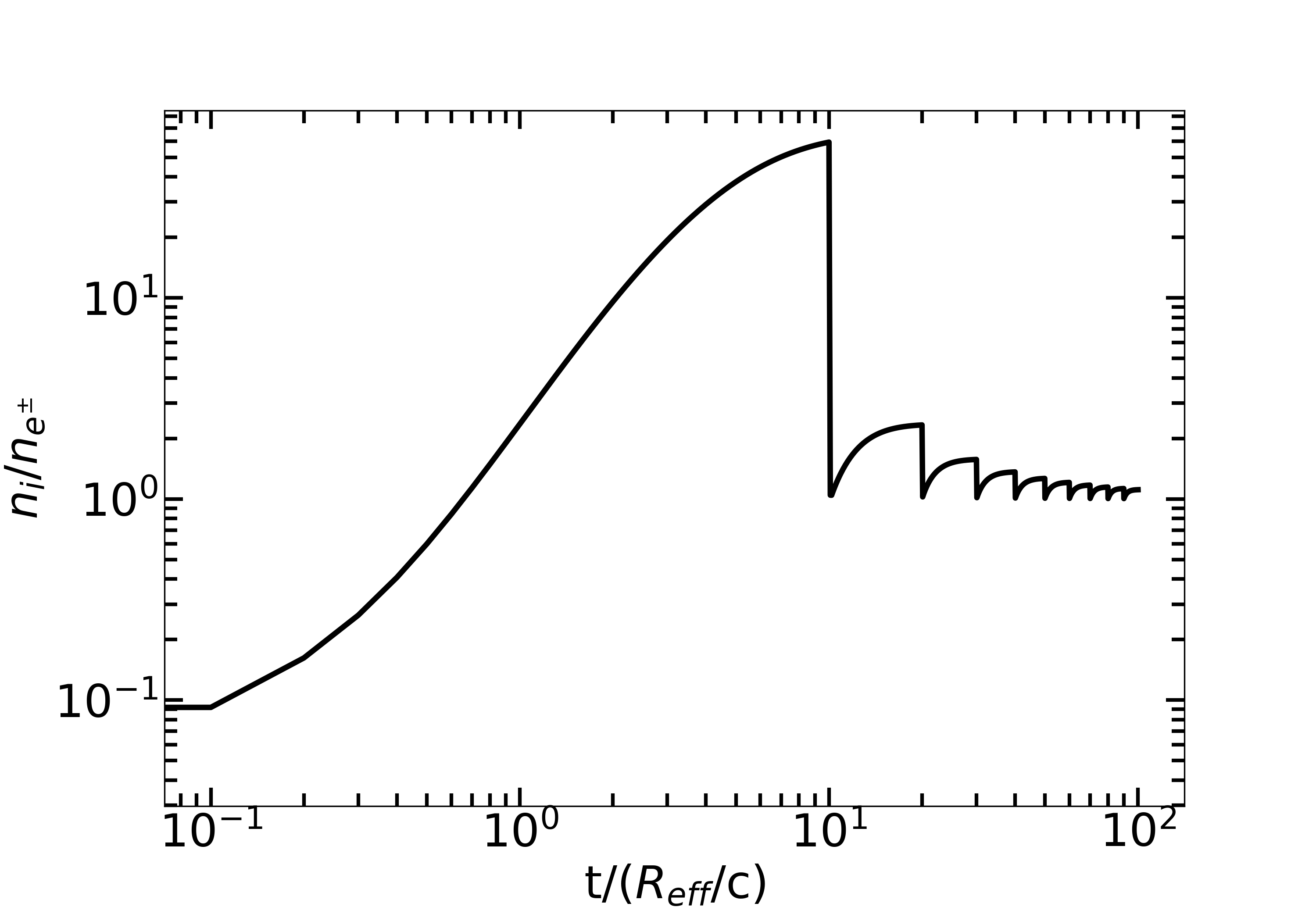}
\caption{ \textit{Left panel:} Temporal evolution of the number densities of three distinct pair populations (``free'', ``trapped'', and ``secondary'') in a system initialized with $\sigma_{\rm e}=10^6$ and $\dot{m}=10^{-6}$. \textit{Right panel:} Temporal evolution of the ratio between the total number density of pairs $n_{\rm i}$ (including secondary particles from $\gamma \gamma$ creation) and the number density of primaries.}
\label{fig:pair_enrichvstime}%
\end{figure*}

We then performed a numerical experiment in which we initialized our system with $\sigma_{\rm e}=10^6$ and $\dot{m}=10^{-6}$. Within the first ten light crossing times, the number density of secondary pairs increases faster with time than the density of primary (free and trapped) pairs in the system, as shown in the left panel of Fig. \ref{fig:pair_enrichvstime}. As a result, the ratio of the total number of pairs in the system to the density of pairs inferred by the initial $\sigma_{\rm e}$ value exceeds unity (right panel in  Fig.~\ref{fig:pair_enrichvstime}), leading to a reduction in $\sigma_{\rm e}$. From this point on, the number density of secondary pairs in the system does not evolve with time, because the pair production rate is independent of $\sigma_{\rm e}$ (see Eq.~\ref{eq:L_sec}). The variation of $\sigma_{\rm e}$ continues until $10^2 R_{\rm eff}/c $ where we terminate the numerical calculation. The density ratio between the total number density of pairs $n_{\rm i}$ and the number density of primaries $n_{\rm e^{\pm}}$ eventually is close to unity (see right panel of Fig. \ref{fig:pair_enrichvstime}) on timescales $\sim 10^2 R_{\rm eff}/c$ that are close to the typical lifetime of the longest current sheets formed in GRMHD simulations. 

\begin{figure*}%
    \centering
\includegraphics[width=0.49 \textwidth]{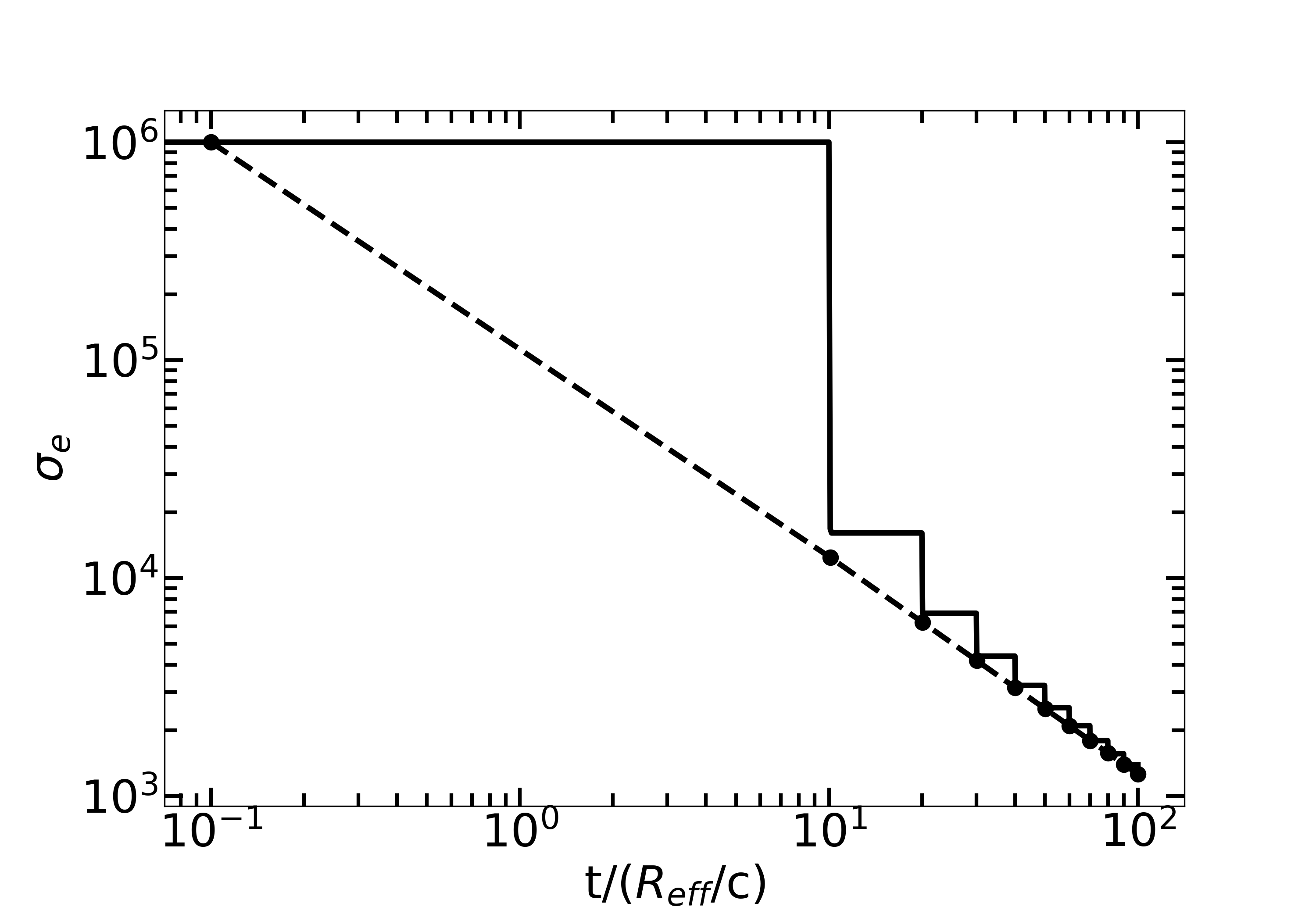}     
\includegraphics[width=0.49 \textwidth]{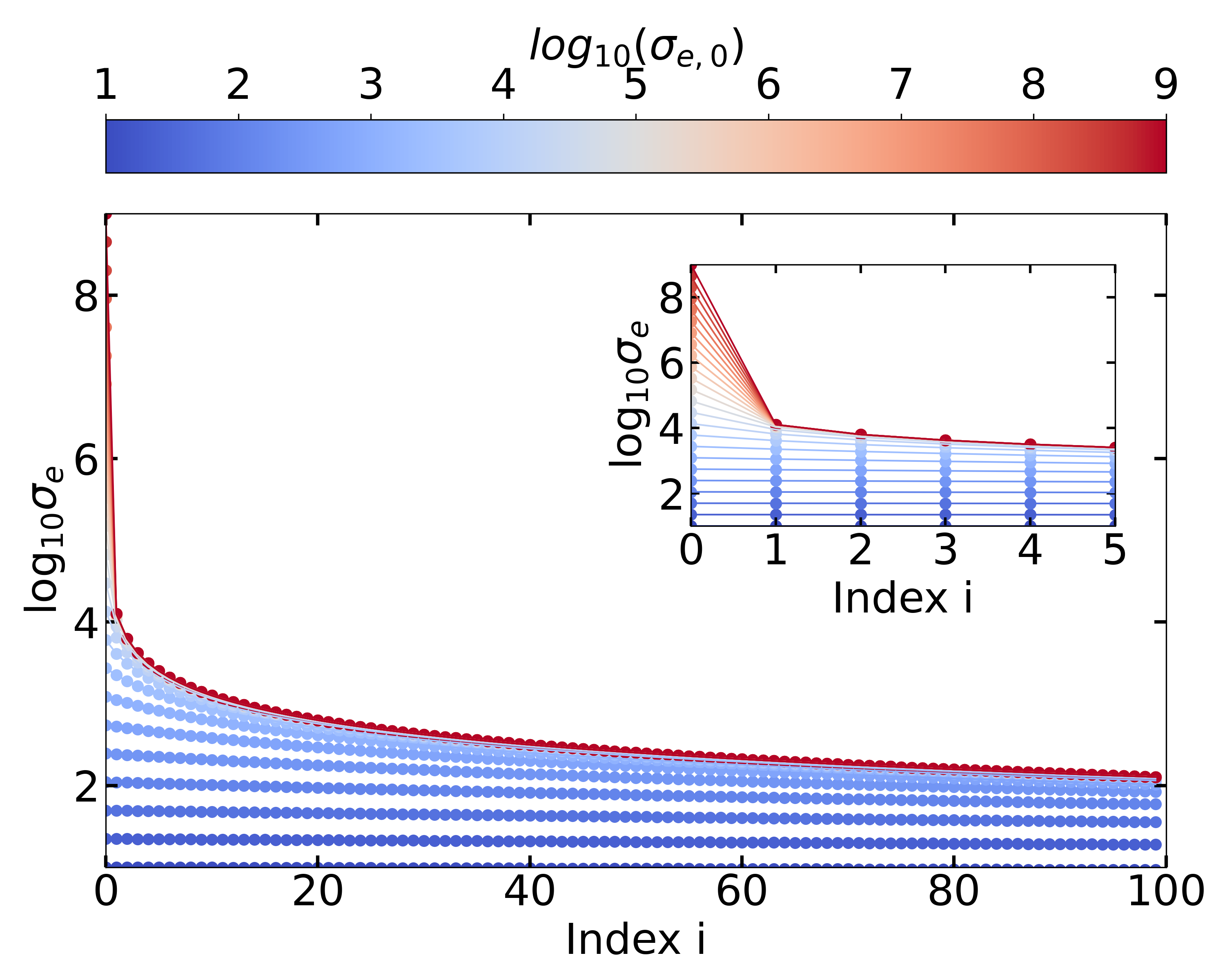}     

\caption{Self regulation of magnetization $\sigma_{\rm e}$ with respect to time.\textit{Left panel:} Evolution of $\sigma_{\rm e}$ for a case with initial magnetization of $\sigma_{\rm e}=10^6$ and $\dot{m}=10^{-6}$. \textit{Right panel:} Evolution of $\sigma_{\rm e}$ for various initial values as represented by the color bar.}
\label{fig:sigma_evol}%
\end{figure*}

In the left panel of Fig. \ref{fig:sigma_evol} we present the evolution of $\sigma_{\rm e}$ with the solid line. As explained above, this step-wise evolution occurs because we adjust $\sigma_{\rm e}$ at intervals of $10 R_{\rm eff}/c$, leading to a period where the magnetization stays constant, followed by a decrease that correlates with the increase of the total pair density in the system in response to secondary pair creation. The black markers represent the analytical expectation of $\sigma_{\rm e}$ when we use the results from Eq.~(\ref{eq:sigma_e_i}). The plasma magnetization has decreased to $\sim 10^3$ by the end of the numerical calculation. Nonetheless, the largest decrease in $\sigma_{\rm e}$ (factor of $100$ occurs at earlier times) pair creation could therefore substantially modify the initial plasma magnetization, even for shorter-lived current sheets. We generalize the results on $\sigma_{\rm e}$ evolution in the right panel of Fig. \ref{fig:sigma_evol}. This figure demonstrates the evolution of $\sigma_{\rm e}$ when we start from various initial values, $\sigma_{\rm e,0}$, as represented by the color bar. The evolution of the system is computed using Eq. \ref{eq:sigma_e_i} for discrete steps i ranging from 1 to 100. We observe that upstream plasma magnetizations $\sigma_{\rm e,0} \gtrsim 10^3$ are going to be regulated because of pair creation after $\sim10$ iterations while $\sigma_{\rm e,0} \lesssim 10^3$ magnetization remains constant.

\end{document}